\documentclass[12pt,preprint]{aastex}

\def\etal{{\it et al.}~}

\def\angst{\mbox{$\,${\rm \AA}}}
\def\kms{\mbox{\,km\,s$^{-1}$}}
\def\degrees{\mbox{$^{\circ}$}}  

\shorttitle{Probing Galaxies Halos using Background QSO}
\shortauthors{C\^ot\'e et al.}

\begin{document}

\title{Probing Halos of Galaxies at Very Large Radii using Background QSOs\footnote{Based on observations with the NASA/ESA Hubble Space Telescope, obtained at the Space Telescope Science
Institute, which is operated by AURA, Inc. under NASA contract
NAS5-26555.}}

\author{S. C\^ot{\'e}} 
\affil{Canadian Gemini Office, Herzberg Institute of Astrophysics, National Reseach Council 
 of Canada, 5071 West Saanich Road, Victoria, B.C., Canada, V9E 2E7} 
\email{Stephanie.Cote@nrc.ca}   

\author{R. F. G. Wyse}
\affil{Department of Physics and Astronomy, Johns Hopkins University,
3400 N.Charles Street, Baltimore, MD 21218, USA}
\email{wyse@pha.jhu.edu}

\author{C. Carignan}
\affil{Observatoire du Mont M\'egantic et D\'epartement de Physique,
Universit\'e de Montr\'eal, CP 6128, Succ. centre ville, Montr\'eal, H3C 3J7, Canada}
\email{carignan@astro.umontreal.ca}

\author{K. C. Freeman} 
\affil{Research School of Astronomy and Astrophysics, Mount Stromlo Observatory, ANU,
Weston Creek ACT 2611, Australia}
\email{kcf@msokcf.anu.edu.au}

\author{T. Broadhurst} 
\affil{School of Physics and Astronomy, Tel-Aviv University, Tel-Aviv 69978, Israel}
\email{tjb@wise.tau.ac.il}

\begin{abstract}
Gaseous halos of nine nearby galaxies (with redshifts cz $<$ 6000
\kms ) were probed at large galactocentric radii using background
quasars observed with HST GHRS and STIS. The projected quasar-galaxy
separations range from 55 to 387~$h^{-1}_{75}$ kpc.  Ly$\alpha$ absorption lines
were successfully detected in the spectra of five quasars, at
impact parameters of up to $\sim$ 170 $h^{-1}_{75}$~kpc from the center of
the nearby galaxy, and in each case at wavelengths consistent
with the galaxy's redshift.  Our observations include the lowest
redshift Ly$\alpha$ lines detected to date.
HI velocity fields were obtained at the VLA for three of the galaxies in our sample (and in one case
was available from the literature), to derive their rotation curves. When comparing
the inner rotation curves of the galaxies with the velocity at large radius provided
by the Ly$\alpha$ line it is apparent that it is very difficult to explain the observed 
Ly$\alpha$ velocity as due to gas in an extended rotating disk.
In most cases one would need to invoke large warps in the outer gas disks and also thick gas disks in order 
to reconcile the observed velocities with the predicted ones. Indeed,
in one case the Ly$\alpha$ line velocity indicates in fact counter-rotation with 
respect to the inner disk rotation.
In light of these results we conclude  that in a typical galaxy there is no longer detectable atomic gas corotating
in an extended  disk at radii $> 35$ $\alpha ^{-1}$, where $\alpha^{-1}$ is the stellar disk exponential scale-length.
The cosmic web is the most likely origin for the detected Ly$\alpha$ lines.
Our observations 
confirm  the Bowen \etal (2002) correlation of 
 equivalent widths  with the local volume density
of galaxies around the sightline, and the observed equivalent widths of the lines are 
consistent with expectations of the cosmic web. 

\end{abstract}

\keywords{galaxies: halos --- galaxies: spiral --- quasars: absorption lines --- galaxies: ISM}

\section{Introduction}

Rotation curves are the best tool to study the dark matter halos of
galaxies, and the most extended ones will better constrain the dark
matter halo parameters. Rotation curves derived from neutral hydrogen
(HI) observations enable us to probe the dark matter potential to
distances of at most 20 to 25 $\alpha^{-1}$ (where $\alpha^{-1}$ is the
optical disk exponential scale length), 
for gas-rich late-type spiral galaxies with particularly
extended HI (eg. NGC2841 from Begeman 1987 or NGC2915 from Meurer
\etal ~1996). But it is not possible to obtain any kinematical data
with 21cm emission further out than the radius at which the gas
density falls to levels of a few times $10^{19}$ atoms cm$^{-2}$,
even with very sensitive observations, as the HI column falls off
sharply beyond this point (eg. the 'HI Edge' observed in NGC3198,
van Gorkom 1991). This sharp truncation of the HI distribution is
believed to be the result of ionization of the atomic gas by the
extragalactic UV radiation field (as predicted by Silk \& Sunyaev
1976). At lower
column densities the neutral fraction is therefore predicted to drop
off dramatically in comparison to the total (ionized) hydrogen column
(see Maloney 1993).  Some studies have attempted to detect the faint
ionized gas just beyond the HI limit from recombination emission in
H$\alpha $. Depending on the volume density of the plasma at the
critical column density where the outer disk becomes optically thin to
this radiation (about $N_{HI} \sim 3\times 10^{19}$~cm$^{-2}$ for
NGC3198), the emission measure from recombination radiation expected
to be emitted by this ionized gas is only 0.025 to
0.25~cm$^{-6}$pc (Maloney 1993).  The only reported detection of
recombination radiation so far is that of NGC253 for which
sensitive Fabry-Perot observations by Bland-Hawthorn \etal ~(1997)
managed to extend the HI rotation curve from 1.2 to 1.4 $R_{25}$, with
measured surface brightness values in H$\alpha$ from 80 to 40 mR (millirayleighs,
where 1 rayleigh = 10$^6$/4$\pi$ photons cm$^{-2}$s$^{-1}$sr$^{-1}$ or 2.78 cm$^{-6}$pc
at H$\alpha$).
But at this level, 
as pointed out by the authors, 
the emission is probably not due to recombination after ionization
by the metagalactic background, but rather from hot young stars near
the center of the galaxy ionizing the warped outer HI disk. 

If indeed spiral galaxies have extended (mostly ionized) gas disks it
should then be possible to detect them in absorption in the spectrum
of a bright background quasar, down to HI column densities of only
$\sim $10$^{13}$ atoms cm$^{-2}$. With such a suitably placed
UV-bright background object one could obtain kinematical information
on that galaxy's outer rotation curve out to 30 to 40 $\alpha ^{-1}$
or more.  At such radii the data could potentially reveal the extent
of the dark matter halo, and constrain the mass profile out to such
large galactocentric radii.  This could then provide strong tests of
cosmological galaxy formation scenarios.  For example, according
to the rotation curves predicted by Cold Dark Matter-dominated
N-body simulations (e.g.~Navarro, Frenk \& White 1996) in low-mass
dwarf galaxies we should detect a turn-down in the rotation curve
if we could get data points just 15 or 20\% further out in radius
than currently mapped by HI observations.  This matching of a
Ly$\alpha$ absorption line velocity with the kinematics of the nearest
galaxy has already been attempted, originally by Barcons \etal (1995),
who found that for two galaxies at z=0.075 and 0.09 the Ly$\alpha$
velocities as measured from background QSOs spectra at impact
parameters 64 $h^{-1}_{75}$kpc (= 16$\alpha ^{-1}$) and 83 $h^{-1}_{75}$kpc (=35$\alpha ^{-1}$),
were consistent with gas in extended gaseous haloes co-rotating with
the inner stellar disks (with $h^{-1}_{75} = H_0/75$ where $H_0$ is the Hubble
constant and $q_0=0$). On the other hand, Hoffman \etal (1998)
found that the dwarf galaxy MCG+00-32-16, which is the closest galaxy
to the low redshift Ly$\alpha$ system in the sightline of 3C273, has an
HI disk which would be counter-rotating compared to the Ly$\alpha$
velocity. In this case however the impact parameter is 204 kpc which
for the dwarf galaxy represents about 153 $\alpha ^{-1}$, in other
words an excessively large radius to expect the extended gas halo to
reach. More recently Steidel \etal (2002) have obtained long-slit
spectra of 5 intermediate redshift galaxies ( 0.44 $\leq$
z$\leq$0.66) in the proximity of MgII absorbers sightlines (with projected 
separations from 19 to 95 $h^{-1}_{75}$kpc), and succeeded in reasonably matching
the MgII absorbers' velocities with the galaxies' rotation curves, but
only when setting various disk thicknesses and velocity scale heights
(describing the velocity fall-off with z-height above the plane) case
by case.
 
Our approach here is the reverse. Instead of identifying and analysing
the galaxies found near the line-of-sight of a quasar exhibiting a
known Ly$\alpha$ line, we start by selecting nearby galaxies, normal
and well-behaved kinematically, hence easy to subsequently model
dynamically, and search for bright background quasars at some
particular impact parameter away, ie: at the interesting radii beyond
the galaxy's HI envelope and up to about 50 $\alpha^{-1}$. 
 We then observe these quasars with HST, first to
attempt to detect a Ly$\alpha$ line arising from the
extended galaxy's disk, and then to use this Ly$\alpha$ line velocity to
probe the outer dynamics of the galaxy.

Our study does not aim at elucidating the overall nature of all
Ly$\alpha$ forest lines (other studies are better designed for this,
by surveying all lines along several QSO sightlines and getting
numerous spectra of galaxies in the surrounding fields eg. Penton
\etal 2002, Morris \etal 2002).  Indeed, while it is now generally
accepted that most metal-line absorption systems found in QSO's
spectra are associated with gaseous galaxy halos, the situation for
the Ly$\alpha $ absorption systems has been long under debate.  Some
studies claim that (at least the stronger) Ly$\alpha$ lines are
associated to extended halos of galaxies, while most believe that the
majority of the lines occur in the intergalactic medium, tracing the
'cosmic web' (cf. van Gorkom et al.~1996), this intricate network
of filaments and sheets of gas predicted by hydrodynamical simulations
(eg. Dav{\'e} \etal 1999).  Some
convincing arguments of the latter are the fact that several
Ly$\alpha$ lines have been detected in voids (McLin \etal 2002), also
confirmed by deep HI observations that would have detected dwarf
galaxies and other low surface brigthness galaxies (Shull \etal
1998). Also observations of double QSOs sightlines find Ly$\alpha$
lines in common in both spectra on scales of typically half a Mpc
(Dinshaw
\etal 1997, 1998).                      
Here we are testing the simple conjecture that gaseous galactic disks extend further beyond that which 
can be probed in emission in HI, by searching for 
 a corresponding Ly$\alpha$ line.  More importantly, a successful detection, 
can be used as a dynamical probe of the outer halo of the galaxy.  

In Section 2 we explain the selection of our targets, section 3 describes the
observations (HST and radio), then section 4 discuss the results of matching
the Ly$\alpha$ velocities with the galaxies' rotation curves and the implications.

\section{Selection of Targets}

Our list of QSO-galaxy pairs suitable for our purpose was compiled by
cross-correlating all 646 quasars with V$\leq $ 17.0 in the V\'eron \&
V\'eron catalogue 6th edition (1993), with the {\it HI Catalog of Galaxies}
(Hutchmeier \& Richter 1989). This magnitude limit was imposed to make sure
that the background QSO observations would stay within a reasonable amount of time
to be feasible with HST. Our criteria for selection were:

{\bf i)} the projected distance between the QSO and the galaxy was
less than roughly 10 times the optical diameter (D$_{25}$) of the galaxy.
This was to find QSOs in the interesting region beyond the galaxy's HI
emission extent but where the column density of neutral gas would
still be high enough to be detected in absorption (assuming that
galaxies have extended gaseous disks out to large radii as suggested by
eg. Chen \etal 2001). 

{\bf ii)} the galaxy systemic velocity was $\geq $ 550 \kms . This
ensured that the associated Ly$\alpha$ detection stayed clear from the
Geocoronal Lyman-$\alpha$ emission at 1216 \angst ~(from the Earth's
exosphere).  Moreover this also ensured that the Ly$\alpha$ detection's
velocity would not fall in the range covered by the 
damped Ly$\alpha$
wing of our Galaxy, which can extend to $\simeq $ 1220 \angst ~towards
some sightlines, depending on the total HI column density present in
that direction. If a galaxy's systemic velocity is too low, any
associated line would get lost in the wing of the Galactic damped
line, where the continuum against which to detect this absorption line
is seriously depleted.

{\bf iii)} the galaxy was not interacting and was relatively
isolated. This means that an observed Ly$\alpha $ line would be fairly
unambigously assigned to that object, ie: QSO-galaxy pairs in the
middle of clusters, such as for 3C273, were discarded
(but one target, NGC5033, was later found to be  
surrounded by several dwarf galaxies, see Section 4.1).

{\bf iv)} the galaxy's optical diameter was at least $\geq 1^\prime$ so
that a detailed kinematical study, using HI rotation curves, 
 is feasible (since the HI radio beams will be in the best cases about 
a dozen of arcsecs). In some cases the rotation curves were already available 
in the literature.  For the same reason the galaxy also had to appear
\lq\lq well-behaved", with no obvious distortions or
morphological asymmetries -- in other words as normal as
possible.

We observed a sample of 9 QSO-galaxy pairs with HST,
chosen to cover a range of morphological types (Sa to Sd), with a range
of magnitudes $-20 < M_B < -15.1$. Assuming successful detections of
Ly$\alpha$ absorption lines, this would allow testing of scaling
relations across a wide range of galaxy properties.  For example Chen
\etal (1998), in their study of 26 galaxy-absorber pairs with
separations ranging from 16 to 209 $h^{-1}_{75}$ kpc, find that the
strength of the absorption depends not only on the impact parameter
separation but also on the B-band luminosity of the galaxy. Hence
galaxies with a wide range of properties were chosen for our sample to
investigate these possible trends. Table~\ref{tab3} lists the
properties of our selected target galaxies. Note that  
the impact parameters of the galaxy-QSOs are based on distances estimated
using $H_0 = 75$ km/s/Mpc. In the case of NGC2841 there is a
Cepheid distance of 14.1 $\pm$ 1.5 Mpc (Macri \etal 2001).  
Its velocity simply estimated with $H_0=75$ would give a distance of 8.5 Mpc,
which is only 60\% of this 'true' distance. Hence this gives an idea of the
possible large uncertainties in these impact parameters estimates.

\section{Observations and Reductions}

\subsection{HST Observations of QSOs}

Spectroscopy of our 9 target QSOs was obtained with the HST GHRS
through the Small Science Aperture (SSA) with the G140L grating, as
well as with the HST STIS/FUV-MAMA through the 52X0.1 aperture
with the G140L and G140M gratings, as listed in
Table~\ref{tab2}. This Table also gives details of the observations
(the HST dataset filenames, date of the observations) and the total
integration times of the spectra. Since the QSOs have a range of
fluxes, the integrations were set such as to result in a 3$\sigma $ limiting
equivalent width of at least 0.3 \AA, or in the case of the STIS data, to
fill completely the one orbit necessary (this 3$\sigma $ limit being in almost all cases
achievable in less than one orbit). 

The preliminary GHRS data reduction was carried out with the standard CALHRS software
using the final GHRS reference files,
and further reduction and analysis were performed with the IRAF/STSDAS package.
To fully sample the PSF, quarter-diode substepping was employed, producing four
spectra per exposure. With this mode 6\% of the time is spent measuring the
background with the science diodes. 
The wavelength shifts between these different groups, and then between the
different exposures, were determined to align all the spectra before combining them
and merging the wavelength and flux information.
With the G140L grating the dispersion is 0.57 \AA ~per diode, corresponding to a 
velocity resolution of about 140 \kms ~at 1200 \AA.  
The default wavelength scale is expected to have a maximum rms of 55 m\AA ~for G140L.
However, zero-point shifts can be significant: even though we used the small
aperture which will limit the effect of the uncertainty in the position of the target
 within the aperture,  thermal effects in the GHRS as well as geomagnetic effects are 
a large source of wavelength error (see GHRS Instrument Handbook). The wavelength
accuracy achievable is about 30 \kms, even when a wavelength calibration exposure
is obtained just before the science exposure. For this reason 
these zero-points were corrected by assuming that low-ionization Galactic 
interstellar lines lie at the same velocity as the observed dominant component of 
the Galactic HI along these sightlines. The strongest Galactic lines were used such as  
Si III $\lambda $1206.5, Si II $\lambda $1260.4 and CII $\lambda $1334.5. The Galactic 
HI LSR velocity measurements 
were taken from the Leiden/Dwingeloo survey (Hartmann \& Burton 1997).

For the STIS observations, the G140M centered at 1222 \AA ~provides a
resolution of 20 \kms ~(with a pixel size of 0.05 \AA) and wavelength
coverage from 1195 to 1249 \AA.  In addition short G140L observations
were obtained for the 3 QSOs at higher redshifts (PG1049-005,
PKS1103-006, and PG1259+593). This was to make sure that any
absorption feature detected at 1218-1222 \AA ~(where our Ly$\alpha$
lines are expected) is not in fact a Ly$\beta$ line associated with
some Ly$\alpha$ line at higher redshift along the sightline. The G140L
therefore provided us with a spectrum over a larger wavelength range
(roughly 1130 to 1720 \AA) to check on this possibility.  The standard
CALSTIS procedures were followed to flatfield the individual spectra,
extract them and then wavelength-calibrate and flux-calibrate
them.  The MAMA spectroscopic accuracy is expected to be about 0.2
pixel when using the narrow slit ($<$3~\kms ~with G140M), but is
ill-defined in practice.  Hence the zero-points were corrected the
same way as for the GHRS data, by shifting Galactic lines in the
spectra to the main Galactic HI LSR velocity. In this case the
Galactic lines used were Si III $\lambda $1206.5 and NI $\lambda $1199.55, 
$\lambda $1200.22, $\lambda $1200.71
when suitable.  For the Galactic HI velocities, several sightlines had
higher resolution HI Galactic spectra from Effelsberg (Wakker \etal
2001, 9'.1 beam), and for the rest the measurements were extracted from the
Leiden/Dwingeloo survey (Hartmann \& Burton 1997, 36' beam), or from
NRAO 43m data (Lockman \& Savage 1995, 21' beam). Care was taken with
galactic features at non-LSR velocities due to High-Velocity Clouds
(HVCs) present in some of our sightlines.  For example the MRK876
sightline crosses the region of Complex C, a prominent HVC covering
~10\% of the northern sky, and consequently the Galactic and Complex C
Si III line are blended in the spectrum (Gibson \etal 2001). Our final
shifts applied to the spectra differed from the original calibration
by 0.8 \kms ~up to 11.6 \kms ~at most.  Finally, after continuum
fitting and normalization with a low-order polynomial, wavelengths and
equivalent widths were measured, and Voigt profiles were fitted to the
observed absorption lines. These profile fits are depicted in Figure~\ref{fig3}.
In the case of PG1309+355, slightly better residuals were obtained if fitting
3 subcomponents to the line, and similarly for PG0804+761 with 2 subcomponents.
For sightlines with non-detections, equivalent width limits
were calculated following the equations of Ebbets (1995).

Ly$\alpha $ absorption lines were successfully detected for 5 of our 
QSO-galaxy targets (illustrated in Figure 1). Figure~\ref{fig2} shows the (unbinned) spectra with
the detections, as well as the 4 spectra with non-detections.
Table~\ref{tab3} lists the detections, and their corresponding velocities and equivalent
widths. In many cases other Ly$\alpha$ lines were also detected along the same sightlines at
other various redshifts, and do not necessarily seem to be associated with any nearby
galaxy. These will be discussed more fully in a forthcoming paper.
In all cases the velocities of our detections agreed
within errors with those in the literature when available, for example from higher resolution FUSE data 
detecting the Ly$\beta$ counterparts of some of our lines. For PG0804+761 our line
of interest at $V_{Ly\alpha}$=1570\kms ~has not been previously reported, but a 
higher redshift line in the spectrum which we detect at $V_{LSR}$=5561 $\pm$ 8\kms ~has
been reported by Shull \etal (2000) at $V_{LSR}$=5565 \kms ~and by Richter \etal (2001)
at $V_{LSR}$=5553 \kms ~(from a FUSE Ly$\beta$ line). For MRK876 Shull \etal (2000)
report a $V_{LSR}$=958 \kms ~(no errors quoted) for which we have $V_{LSR}$=950 $\pm 5$ \kms.

For the equivalent widths, a large part of the uncertainties come from the continuum fitting,
especially since some of the lines are sitting on the edge of the Galaxy's Ly$\alpha$ 
damped wing. Hence the errors as presented in Table~\ref{tab3} are the sums in quadrature of the uncertainties
(1$\sigma$) from photon noise and the estimated errors from different reasonable placements
of the continuum fit. The continuum Signal-to-Noise per resolution element in our spectra
in the region of the expected or detected absorption ranges from 2.2 to 12,  
since our target QSOs have a wide range of fluxes (the lowest S/N of 2.2 is for 
PKS1103-009 which had a detected flux of only 3/5 of the value estimated from IUE data).

\subsection{HI Observations and rotation curves}

To compare the five detections of Ly$\alpha$ absorptions in spectra of background QSOs
with the associated galaxy's kinematics we needed its velocity field and rotation curve.
For NGC5033 and UGC8146 these were available in the litterature (Begeman 1987; Rhee \& 
van Albada 1996).
For the other 3 galaxies we carried out Very Large Array (VLA) HI observations,
for which the observational parameters are summarized in Table~\ref{tab4}.
The correlator was set to a 3.125 MHz bandwidth with 128 channels,  
 for a channel separation of 5.2 \kms. The calibration
and reductions were performed with the NRAO package AIPS, following standard 
procedures. The absolute flux calibration was determined by observing the standard source 
1331+305. After applying calibration and bandpass corrections to the {\it uv}
databases, the continuum emission was subtracted in the visibility domain 
using channels free of line emission. Several sets of maps were produced 
for each galaxy, one with uniform weighting preserving the best resolution,
and a few with natural weigthing smoothed to various larger beams. These sets
of maps were CLEANed well into the noise by monitoring the total cleaned 
flux as a function of the total number of CLEAN components recovered, and
finally were restored with a circular symmetric Gaussian beam. These maps 
were then corrected for the beam response of the antennae to rectify for 
attenuation away from the center of the primary beam. No HI emission was
detected from other galaxies within the primary beam and through the bandwidth
observed other than the intended source.

Moment maps were obtained from each datacube using Hanning smoothing in
velocity and Gaussian smoothing spatially, to produce integrated total 
HI column density maps, intensity-weighted velocity maps, and velocity
dispersion maps.
HI rotation curves were obtained by fitting a tilted-ring model to the velocity fields
(Begeman 1987, C\^ot\'e \etal 2000), and are given in Table~\ref{tab5}. 
In each case the rotation curves are slowly rising and just barely reach
the flat part at the last measured points.
The errors in velocity were 
calculated from half the difference between the velocities on each side (receding
and approaching) or from
the formal errors given by the least-squares fits, whichever were the highest.
The rotation curves were derived with the systemic velocities and orientation
parameters (position angle and inclination) set to the values shown in 
Table~\ref{tab4}. The derived systemic 
velocities agree extremely well with the RC3 
values, within 1 \kms. The orientation parameters as well agree within a few
degrees with those derived optically (from the RC3), which show that these 
galaxies
do not have strong warps in their outer HI gaseous envelopes. The HI distributions
and HI velocity fields are shown in Figure~\ref{fig4}, and the gas 
suface density profiles in Figure~\ref{fig5}, for which the HI surface
density values were scaled by 4/3 to account for the presence of 
helium. Although the three surface density profiles differ greatly,
they are all within the range of what is commonly detected in late-type
spirals.

\section{Results}

\subsection{Ly$\alpha$ detections}

Figure~\ref{fig2} show the 5 Ly$\alpha $ absorption detections amongst
our 9 target QSOs, and Table~\ref{tab3} gives their details.  
The target galaxy-QSOs which give rise successfully to a Ly$\alpha $
detections have impact parameters from 55 kpc to 169 $h^{-1}_{75}$~kpc, and the detections
correspond to HI column densities ranging from log N$_{HI}$=13.0 to 13.9 cm$^{-2}$. 
PG1309+355-NGC5033
with an impact parameter of 276 $h^{-1}_{75}$~kpc, also give rise to a detection, 
however in this case it was later found
that NGC5033 is part of a small loose group of galaxies and is surrounded
by a swarm of dwarf galaxies, some of which are closer to the QSO sightline 
than is NGC5033  (eg., UGC8261, UGC8323, UGC8314); and it is hence not possible 
to attribute the absorption 
detection to NGC5033 alone. We will return to this case later on.

In each case the Ly$\alpha $ line velocity matches extremely well the
systemic center-of-mass velocities of the galaxies. In particular 
for PG 1259+593 and MRK 876 their Ly$\alpha $ absorption
lines correspond to velocities of 679 and 935 \kms
~(respectively), which are the lowest redshift Ly$\alpha $ lines ever
detected so far in association with nearby galaxies. Presumably
closer galaxies,  e.g. M31, could produce Ly$\alpha $ lines of
even lower redshift but they would be impractical to detect because of
the damped wing of the Milky Way Galaxy absorbing away all the continuum flux.
In fact in Figure~\ref{fig2}, in all our spectra, the slope of the Galaxy's damped
wing is clearly seen, ie: the Ly$\alpha $ detections are sitting inside
the range covered by the Galaxy's damped wing which extends to at least
$\sim$ 1220 \AA, sometimes 1222 \AA.

As for the Ly$\alpha$ non-detections, they occur for galaxy-QSO
separations ranging from 162 to 387 $h^{-1}_{75}$~kpc. Their 3$\sigma$ limiting
equivalent width is in each case above the level needed to detect an
absorption line even weaker than the ones detected in the other five
spectra. In terms of HI column density limits it corresponds to a
3$\sigma$ detection limit around log N$_{HI}$=13.1 cm$^{-2}$ on
average. This at first sight thus appears to confirm the findings of
several studies of moderately low redshift galaxy surveys
associated with absorbers, which seem to point to most galaxies being
surrounded by tenuous gas out to $\sim $ 240 $h^{-1}_{75}$ kpc. Chen \etal
(1998) and (2001), in a continuation of the work of Lanzetta \etal
(1995), find for 34 galaxies and absorbers pairs at z=0.07 to 0.89
with impact parameters $\rho$ from roughly 16 to 233 $h^{-1}_{75}$ kpc, a
near unity covering factor for tenuous gas halos around typical L$_*$
galaxies out to a radius of $\sim$ 240 $h^{-1}_{75}$ kpc.  Furthermore Chen
\etal (1998) found that the absorbing gas equivalent width depends on
the galaxy-QSO impact parameter as well as the galaxy B-band
luminosity, which if true would greatly strengthen the case for a
direct association between the absorber and its nearest galaxy.
Meanwhile Bowen \etal (1996) found that for 38 galaxies at z=0 to 0.08
lying at $<$ 400$ h^{-1}_{75}$ kpc from a sightline of a bright QSO observable 
with FOS, 
 for lines with equivalent widths $>$ 0.3\AA the
covering factor is 44\% between 67 to 400 $h^{-1}_{75}$ kpc. Recently Bowen
\etal (2002) detected Ly$\alpha $ lines in the outer regions of all
eight nearby galaxies probed, with impact parameters up to 265
$h^{-1}_{75}$ kpc, down to $W>0.1$\AA.  Our results concord with these
studies. All of our detections occur at impact parameter $<$ 180 $h^{-1}_{75}$ kpc
from the nearest galaxy, and only one of our
non-detections occurs for $\rho <$ 180 $h^{-1}_{75}$ kpc (PKS1103-006 - NGC3521 at
$\rho$=162 $h^{-1}_{75}$ kpc). Note however that there is considerable uncertainty
in the values of $\rho$ due to the distance dependence (see
above).  All these results seem very suggestive of a link between the
detected absorbers and their neighboring galaxies. 
 But one needs to
look into the detailed kinematical match between the Ly$\alpha$
absorption line and the kinematics of the inner galaxy to probe more
directly any direct link between an extended gaseous disk or halo of
the galaxy and the absorber's velocity.

\subsection{Matching the galaxies' kinematics with the absorbers}

We will try to compare here the velocity of the QSO Ly$\alpha $ system,
with that of the associated galaxy. For 4 out of 5 detections, the
associated galaxy is confirmed to be isolated, hence the absorber
can be unambiguously matched to this galaxy, assuming it is indeed due to a galaxy. For 3 of these galaxies,
UGC4238, UGC7697 and NGC6140, we obtained the HI rotation curves
(section 3). The 4th one, UGC8146, has been observed by the Westerbork 
Synthesis Radio Telescope and its rotation curve was extracted from 
Rhee \& van Albada (1996).

To compare the inner galaxy dynamics with the velocity measured
by the Ly$\alpha$ line, we will assume that the line does indeed arise 
in the extended gaseous disk of the galaxy and see how well it matches
the velocities of the HI rotation curve in that case. The observed
radial velocity of the line, $V_{obs}$, is related to the corresponding
rotation velocity $V_{rot}$ in the disk of the galaxy by:
\begin{equation}
V_{obs} = V_{sys} + V_{rot} sin(i) cos(\theta )
\end{equation}
where $V_{sys}$ is the systemic velocity of the galaxy, $i$ is its
inclination, 
and
\begin{equation}
cos(\theta ) = {X\over R}
\end{equation}
where the QSO has coordinates, in the plane of the sky, of X along the 
galaxy major axis, and Y along the minor axis, with
\begin{equation}
R = \sqrt{X^2 + Y^2 sec^2(i)}
\end{equation}
R is the radius in the plane of the galaxy at which the measurement lies,
in other words, the deprojected impact parameter of the QSO 
(a figure demonstrating these formulae is provided in Kerr \& de Vaucouleurs
1955, in which however the inclination angle is defined as $i$=0 for an edge-on 
galaxy contrary to the modern convention). For the orientation parameters
of our galaxies we have used those derived from the fits to their HI velocity
fields.

Figure~\ref{fig6} shows the resulting rotation curves.  The inner points
show the velocities obtained from the HI velocity fields, and the single
outer point is from the Ly$\alpha$ detection. The plotted curves are simply
the best fits CDM models (Navarro, Frenk \& White 1996) to the inner 
rotation curve, and serve to guide the eyes to what velocities might be expected
in these outer parts (note that such CDM curves are actually bad fits to 
observed rotation curves in the inner parts, see, e.g. C\^ot\'e \etal 2000,
but here we just need them as rough approximations for the outer parts). 
The rotation curves in every case just reach the flat part at the very last 
measured point(s) (see Table~\ref{tab5}). 
Each case is discussed separately below.

\subsubsection{NGC 6140}
NGC 6140 is the only galaxy for which the Ly$\alpha$ velocity could be
reasonably reconciled with the value expected if indeed arising
from an extended gaseous corotating disk. Although the Ly$\alpha$ line
velocity is much lower than the maximum rotation velocity $V_{max}$ of
the inner rotation curve, or than the velocity expected from CDM
models at that radius, there are some ways to make it consistent with
a disk velocity at that radius.  For example a strong warp in the
orientation parameters of the galaxy in the outer parts would change
the deprojection factors and hence the calculated Ly$\alpha$
rotation velocity could rise to match the expected value. The present Ly$\alpha$
velocity was calculated using the HI orientation parameters.  One
would need a warp as large as about 40\degrees ~in inclination, or
alternatively of 22\degrees ~in inclination and -31\degrees ~in
position angle, to bring the velocity in line with the disk expected
one. Most galaxies are observed to be warped in their outer parts, but
such strong warps are rare for isolated galaxies. But although
unusually large, one cannot a priori dismiss the possibility of such a
warp in NGC6140.  Another possibility to consider is if the disk, and
the dark matter halo as well, of NGC6140 are truncated -for
whatever reasons- just outside the radius of the last measured
point in HI. In this case the velocities in the outer parts should
follow a keplerian fall-off, and the velocity expected at the radius
of the Ly$\alpha$ velocity would fall to about 46.2\kms, close to the
measured $V_{Ly\alpha}$=36.2 $\pm$ 5 \kms.  Finally another
possibility is that the velocity measured is not due to gas
actually in the disk plane, but from material at a certain $z$ height
in a thick disk or halo. Assuming a thick rotating disk in which the
circular velocities decrease rapidly as a function of scale height,
$v(z) = v_d e^{-|z|/h}$, where $v_d$ is the velocity in the disk
midplane and $h$ is a velocity scale height, then one can easily
recover from the model a lower velocity like the one observed here by
choosing an appropriate value of $h$. Except in the case of an
infinitely large $h$ where the thick disk is corotating
cylindrically, the velocities at some $z$ distance will be much
lower than the midplane one This is observed in NGC~891 for example, 
in which some HI halo gas extending up to 5 kpc from the plane
is seen to rotate 25 to 100 \kms more slowly than the gas in the plane
 (Swaters, Sancisi \& van der Hulst 1997). Hence the observed
Ly$\alpha$ velocity of NGC6140 would not be at all unphysical but
would simply reflect the presence 
 of a rotating thick disk (with detectable gas only at higher z height, and
none in the plane of the disk). For the moment, with only one sightline
velocity at large radius, it is not possible to discriminate between
these various possibilities and confirm any of them.

\subsubsection{UGC4238 and UGC8146}

The situation for UGC4238 and UGC8146 is different. In these two cases, assuming planar kinematics, 
the measured Ly$\alpha$ velocity corresponds to a circular velocity actually larger
than the expected CDM one or even the $V_{max}$ reached by the inner 
rotation curve. A very significant increase of rotation velocity with radius 
would be required to explain the observed velocity. However one can again invoke warps to make the observed velocity
consistent with the extrapolation of the inner disk rotation. In fact in
these two cases, because of the particular orientation parameters of the
galaxies and their angle with respect to their associated quasars' 
sightline, one would need only mild warps (of the order of 10\degrees 
~in the suitable direction) to make the match. Consequently one can still not
dismiss a priori the possibility that the absorptions indeed arise 
from co-rotating gaseous extensions of the inner disks, albeit non-planar motions.

\subsubsection{UGC7697}

The QSO, TON1542, is situated at the South-West of the galaxy;
the observed velocity ($V_{obs}=2557$ \kms) is greater than the systemic
velocity ($V_{sys}$=2536); yet the galaxy's approaching side is towards 
the West, meaning the velocities should all be smaller than $V_{sys}$.
The observed velocity from the absorption line is counter-rotating
with respect to the disk velocities. 
And the observed velocities and geometry of the case of UGC7697 are such
that warps cannot provide consistency with an extended gas disk. 
Unusually large warps would be required
to reverse the inclination of the disk on the sky so that counter-rotation
is observed, and such twisted rotating disks would be unlikely to be stable
(with the exception of nearly face-on galaxies where the reversal of the 
inclination would be easier to achieve; see for example Cohen 1979 or
 Appleton, Foster \& Davies 1986, 
where such a possibility is discussed to explain the apparent
velocity reversals in IC10 and M51). Counter-rotation at very large radii
has been observed in HI in a few galaxies, but those are always 
galaxies interacting with a nearby dwarf (for example NGC4449, Hunter 
\etal 1998) or which also harbour inner counter-rotation and hence 
easily stand out morphologically (like NGC4826, the 'Evil-Eye' galaxy,
Braun \etal 1994). Neither of these cases apply here for any of the galaxies
in our sample.
In the case of UGC7697 no amount of disk thickening, or warps could explain
the discrepancy (unless abnormally large). There are simply no physical
way to produce such a velocity with an extended corotating disk.
This is sobering because UGC7697 is one of the galaxies for which the QSO is
at the smallest impact parameter, and yet the absorption's velocity does not
match. It is also interesting to note from Figure 1 that for UGC7697 the QSO
lies close to the projected optical major axis in angle (compared to the other
galaxies). If the gas has a higher column density in the disk than in the halo
we should have expected a clear detection of the rotating HI disk of UGC7697,
if it does reach out to this radius at the $N_{HI}=10^{13} cm^{-2}$ level.

Note that some have invoked the existence of very low-surface brightness
galaxies or dwarf galaxies, undetected in the optical, which could lie closer to the QSO sightline than the apparently 
nearest galaxy and from which the Ly$\alpha$ line
would originate. In the case of UGC7697  
this
is very unlikely since  we also acquired 
deep HI pointings with the VLA 
on all of our QSOs fields and should have detected such galaxies, down to a few 
10$^6 M_{\odot}$ (to be presented in a forthcoming paper).

\subsubsection{From the literature: NGC988 and NGC3942}

Our conclusion that the Ly$\alpha$ absorptions
cannot arise in simple extended disks are confirmed by two
other galaxies associated with Ly$\alpha$ absorbers taken from the
literature, for which similar results are found.  Bowen, Pettini, and
Blades (2002) targeted a sample of eight nearby galaxies near
the sightlines of QSOs and AGNs. Most of their galaxies are not
isolated (contrasting with the careful selection criteria of our
sample), but are part of binary pairs or small groups (their sample
was produced by cross-correlating the RC3 with the Veron-Cetty \& Veron (1996) 
catalog, and choosing galaxies with $>$ 1300 \kms, and with impact parameters
to QSOs of less than 200 $h^{-1}_{100}$ kpc). However two 
galaxies, NGC988 and NGC3942 are reasonably isolated, at least on
the scale of a few hundreds of kpc. Their projected 
separations from their associated QSOs are 211 and 123~$h^{-1}_{75}$ kpc
respectively, and no other galaxies are known around a diameter of 400
kpc from these QSO sightlines.  Unfortunately no rotation curves,
either optical (in H$\alpha$) or HI, are available for these two
galaxies. But for the sake of comparing the inner galaxy's kinematics
with the Ly$\alpha$ absorption velocity, we can use NGC988 and NGC3942's 
magnitude to derive approximate 
rotation curves, using the \lq universal rotation curve' of Persic
\& Salucci (1996) (where they obtained, based on a sample of 1100
rotation curves, a relation between the shape and amplitude of a 
rotation curve, and the luminosity of the galaxy).
We will see below that these approximations are more
than satisfactory to be able to draw our conclusions. The Ly$\alpha$
line velocities observed in the spectra of MRK1048 and PG1149-110
respectively are taken from Bowen \etal (2002). One last important
piece of information needed to compare the inner and outer velocities
is that the sense of rotation of the galaxy must be known, ie. we must
determine if the QSO lies behind the approaching or the receding
side of the galaxy (otherwise it is not known if the velocity
difference is due to corotation or counter-rotation). This might be
difficult to establish without a proper 2D velocity field for these
two galaxies. Fortunately NGC988 and NGC3942 are reasonably
nearby and extended in angular size, and hence their HI envelope
should be at least a few arcmins in diameter. In that case it is
possible to detect the change in shape of the HI global profile at
different positions around the galaxy in the HIPASS scans (HIPASS is
the All-Southern-Sky HI survey performed with the Parkes Multibeam
receiver and available on-line, Barnes \etal 2001). By retrieving the
HI profiles on each side of the galaxy (the pointings in the on-line
catalog are separated by roughly 8 arcmins from each other), it is
possible to follow the change from a double-peak profile for NGC988
(with a systemic velocity of 1504 \kms) to a single peak roughly at
1400 \kms ~at the East to one around 1600 \kms ~at the West, and hence
determine that for NGC988 the approaching side is at the East and the
receding one at the West.  Similarly for NGC3942 the approaching side
is seen to be in the South-East and the receding one in the NorthWest.
With this information one can then plot the rotation curves of
Figure~\ref{fig7}. They confirm the findings on our galaxies above.
Because for NGC988 the velocity difference between the Ly$\alpha$
absorption line and the systemic velocity of the galaxy is already
quite large, this translates into a rotation velocity (with the
appropriate deprojections) which is ridiculously too elevated ($\sim
$1192 \kms), and no amount of warping, no matter how severe, can bring
this velocity down to a value close to V$_{max}$. For NGC3942 the
calculated rotation velocity is also inadequate, as we find here
another case of counter-rotation. 
Again NGC3942 has a relatively
small impact parameter to the Ly$\alpha$ line, compared to those in
our sample, with the projected distance corresponding to 147 $h^{-1}_{75}$ kpc, and
yet already at that radius we do not detect any longer the extended
corotating gaseous disk of the galaxy.

\subsection{The nature of the Ly$\alpha$ detections}

Could it be that somehow, although not in a corotating disk or halo,
this detected gas is still associated to the galaxy? 
After all the Milky Way is known to be surrounded by High-Velocity-Clouds,
which can have velocities above, or below, the Galaxy's rotation velocity by
several hundred \kms ~(see, e.g. Wakker 1991).  Following strong supernovae events some ejected 
gas might hide into a hot phase until eventually it cools down and could
then be detected to low column densities in our sightlines at very
large radii. Multiple supernovae remnants can drive a
superbubble evolving quickly into a blowout, leaving a hole in the galactic
disk (eg. MacLow \& McCray 1988). Some gas can even be accelerated beyond the 
escape velocity and be 'blown-away', completely unbound from the galactic
potential (de Young \& Heckman 1994). For small dwarf galaxies, for which
it is easier to escape the potential, it might be possible to generate
large (100 kpc size) gaseous halos with supernovae-driven winds (Nath \&
Trentham 1997). For larger galaxies it is expected that there should be 
a local circulation of gas (to a few scalelengths above the disk) via 
a fountain-type flow (Shapiro \& Field 1976).  
However it might be unlikely though to blow 
gas out of a more normal galaxy (closer to $L_*$) to distances of more than
100 kpc, corresponding to factors of up to 7.5 times the diameter of the
galaxy -as is the case here for NGC6140.

Another possibility is that we are intercepting satellites objects (very
small galaxies or clouds of gas) orbiting around the primary galaxy. 
This would explain the counter-rotating velocities, as satellite objects
can have retrograde as well as prograde orbits. In fact in the study of
the supermassive disk galaxy NGC5084 orbited by a record number of 9
satellite dwarf galaxies (Carignan \etal 1997), the large majority
of the satellite objects (7 out of 9) are on retrograde orbits.
The problem with this explanation though is the overly large covering factor
of the detected gas for impact parameters $<$ 160 kpc. 
In the Bowen \etal (2002) study they find a detection for {\it all} QSOs at 
impact parameters $<$265 $h^{-1}_{75}$kpc (down to $W>$ 0.11 \AA), and in our case
 all QSOs at $<$ 136 $h^{-1}_{75}$ kpc yield a detection. It would thus be very
unlikely that by chance the sightlines always intercept one of the 
satellites, considering especially the range of orientation parameters of 
the targets. Moreover these satellite objects would have to have very unusual
properties compared to other dwarf galaxies, since we have deep VLA HI pointings
of all our fields in which we would have detected any objects above about 
10$^6$ M$_\odot$
(to be presented elsewhere).

Another important clue in constraining the nature of the detected
Ly$\alpha$ absorbers is the inspection of the lines' equivalent
widths.  A strong argument in favour of the extended galactic disks'
origin of the Ly$\alpha$ lines has been the apparent anti-correlation
between equivalent width and impact parameter for the lines, the
absorption being stronger the closer it is to the galaxy (Lanzetta \etal 1995).
The Chen \etal (1998) study find also a dependence
on the luminosity of the galaxy, of the form:
\begin{equation}
log ({N\over 10^{20} cm^{-2}}) = -5.33 log({\rho \over 10 kpc}) + 2.19 log ({L_B\over L_{B*}}) + 1.09
\end{equation} 
meaning that brighter galaxies would have stronger absorptions at a
given radius. Tripp \etal (1998) discussed a few possible selection
biases that could explain this anti-correlation. Most interestingly
Bowen \etal (2002) plotted this relation for their 8 nearby galaxy-QSO
pairs and found that the Chen \etal luminosity dependence is not
confirmed. In fact they found a correlation in the {\it opposite}
sense, significant at the 2.8$\sigma$ level, implying that stronger
absorbing lines are associated with fainter galaxies. They concluded
that this relation is probably coicidental, simply a result of small
number statistics.  In our data, with four points, we find a dependence 
in-between that of Chen \etal (1998) and Bowen \etal (2002), in fact it is
consistent with no dependence on luminosity. Meanwhile the hydrodynamical simulations of
Dav{\'e} \etal (1999) have been able to succesfully reproduce the
anti-correlation of equivalent widths and impact parameters in the
context of the cosmic web. They showed that the cosmic gas which
loosely trace the large scale structure will have 3 phases: the
strongest absorbers arise near galaxies because the gas is denser
there; the majority of the absorbers will arise at impact parameters
between 40 and 360 $h^{-1}_{75}$kpc from shock-heated gas around galaxies; and at
impact parameters further than 360 $h^{-1}_{75}$kpc the absorbers will be
associated with a cooler diffuse gas component.

This is corroborated by another interesting correlation, plotted by 
Bowen \etal (2002), between the equivalent widths and the local volume density
of galaxies. For each of the sightlines they calculated the total
equivalent width by summing all lines within $\pm$ 500 \kms, and plotted
them against the number of galaxies with $M_{lim} < -17.5$ in a cylinder of 
2 $h^{-1}$ Mpc radius with a length equal to the distance between
+500 \kms ~and -500 \kms ~from the Ly$\alpha$ line. Their choice of $M_{lim}$
was dictated by the limiting apparent magnitude for completeness in the RC3
along their sightlines. They find that the higher the density of 
$M_{B} < -17.5$ galaxies in a given volume, the stronger the equivalent width 
of Ly$\alpha$ absorption over 1000 \kms. This suggests that the strength of the
Ly$\alpha$ absorbers are not related to their nearest galaxy neighbour but rather
to the overall volume density of galaxies within a few Mpc of the sightline.

In Figure~\ref{fig8} is plotted similarly our L$\alpha$ lines versus the estimated
volume density of galaxies around our targets. Following Bowen \etal (2002) we summed
all components of Ly$\alpha$ absorption within +500 \kms ~and -500 \kms ~from the main
 Ly$\alpha$ line. Only in one case (PG0804+761) was there another Ly$\alpha$ absorber
to add up within that range. We considered similarly only galaxies with $M_{lim} < -17.5$ 
within a 2 $h^{-1}$ Mpc radius and within +500 \kms ~and -500 \kms ~in the RC3.

Our data agree very well with the Bowen \etal (2002) result, within errors
(our errors in $n$ are simply $\sqrt n$). The sightline to TON1542 goes through
the least dense area (with only 2 bright enough galaxies within the 2 $h^{-1}$ Mpc radius)
and the Ly$\alpha$ line is indeed the weakest of all our targets. On the other hand,
the sightline to PG1309+355 and close to NGC5033 is surrounded by galaxies (in fact
we had to a posteriori eliminate NGC5033 from the rotation curve analysis precisely 
because of this, there are too many other galaxies close to the QSO and  hence it was not
possible to attribute the Ly$\alpha$ detection to a galaxy in particular). And  
the Ly$\alpha$ line in the spectrum of PG1309+355 is by far the strongest and most complex,
blending several components into a line with total equivalent width of 1.14 $\pm$ 0.09 \AA. 
The other three sightlines, with small equivalent widths, fall in-between the former
and the latter in terms of volume density of galaxies. The three datapoints are all lying
at higher volume density than the Bowen \etal\ points of similar equivalent width, but considering
the small number statistics here it is probably not a significant offset. These galaxies all lie
at smaller systemic velocities than those studied by  Bowen \etal  so it is possible that the neighborhoods of these local
galaxies have been better surveyed. On the other hand these galaxies
are so close ($< 1500$ \kms) that the estimation of the galaxy volume density by adding
all galaxies within +500 \kms ~and -500 \kms ~probably does not make much sense considering
the large deviations from a smooth Hubble Flow that are seen at these small  distances from us. 

While the choice of 2 $h^{-1}$ Mpc radius for estimating the galaxy volume density in Bowen \etal 
(2002) was rather arbitrary, the limiting $M_{B} < -17.5$ was guided by the estimated completeness
to $B_{lim}=15.5$ of the RC3 for the furthest targets of their sample. In our case because
our galaxies are closer we can inspect the volume density down to a fainter limit using
the RC3. In fact we know that our strongest line, for PG1309+355, is surrounded by dwarf
galaxies in close proximity, but which were not counted in the previous volume density
estimate because they were of lower magnitude than the $M_{lim} < -17.5$. For Figure~\ref{fig9}
we have recomputed volume densities of galaxies down to $M_{lim} < -16.5$, which corresponds
to a $B_{lim}=15.5$ for our sample objects, and in a smaller cylinder of 0.5 $h^{-1}$ Mpc radius.
Again the relation holds very well, although we are dealing with even smaller number statistics.

It thus appear that there is a relationship between the Ly$\alpha$ column density along a 
sightline and the surrounding volume density of galaxies. This makes sense if 
the absorbers are simply part of the cosmic web. The galaxies have condensed in the densest
parts of this network of filaments and sheets of gas, and hence the Ly$\alpha$ detections
will be denser when crossing a web filament/sheet rich in galaxies.
The Ly$\alpha $ absorbers are only detected within about $<170$ $h^{-1}_{75}$~kpc down to our limiting
equivalent widths in the neighbourhood of galaxies because that is where the gas has 
higher column density in the filament, in the dense region from which the galaxy has
condensed. On a larger scale, the Ly$\alpha $ line will be stronger when crossing the core
of a filament, or a busy 'intersection' of filaments of the web, where one finds a higher 
density of galaxy, on a scale of 2 $h^{-1}$ Mpc radius.
That the relationship seems to still hold  on the smaller scale of 0.5 $h^{-1}$ Mpc radius
when considering the density of lower luminosity galaxies (Figure~\ref{fig9}) is perhaps
not surprising, as small groups of galaxies (like the agglomeration around PG1309+355) 
tend to be found on the periphery of larger clusters while most of our other targets
are probably in semi-void regions (since galaxies were purposely selected to be isolated for
the aim of this project). It is not obvious what should be the relevant scale to calculate
the surrounding galaxy density. In simulations of the cosmic web, the filaments come in
various thickness although are definitly of the order of the Mpc, down to the
detectable column densities as here (Miralda-Escud\'e \etal 1996, Dav{\'e} \etal 1999).  
Observations of double QSOs sightlines also find that, by looking at Ly$\alpha$
lines in common in both spectra, the typical scale is around 0.5 Mpc (Dinshaw 
\etal 1997, 1998).

It is interesting to look also at the Doppler $b$ parameters from the
Voigt profile fits and compared them to those obtained by Bowen \etal
(2002) (keeping in mind that some $b$ values are uncertain because of
the $W-b$ degeneracy for saturated lines). In most of their sightlines
several resolved Ly$\alpha$ lines were added up (within their +500 and
-500 \kms ~limit) to add up to the total equivalent widths plotted in
Figure~\ref{fig8}. In constrast, for most of our sample, the total
equivalent widths come mostly from a single line. PG0804+761 is an
exception, for which two lines were added, and also because the main
line associated with the galaxy (ie. which matches more closely in
velocity) is better fitted by two profiles rather than
one. Similarly for PG1309+355, the profile, which has the highest
equivalent width of all, is very broad and show several features,
and the fit is considerably improved by fitting 3 voigt profiles
rather than a single one. Interestingly the trend of
Figure~\ref{fig8} holds well whether one is considering the
addition of several narrow lines (typically $b<50$\kms) in the case of
the Bowen \etal (2002) sample, or the denser, wider single
$b$ lines from our sample. One cannot exclude that our single lines
are actually unresolved groups of narrower lines. If not there might
be a real physical difference in the density and temperature 
of the cosmic web gas depending on the environment, since
our galaxies were mostly selected to be isolated, while the Bowen
\etal (2002) ones are part of pairs or small groups. The Bowen \etal
(2002) lines are also at redshifts on average twice as distant 
than for our sample.  While some evolution with redshift is expected
for the density of the cosmic web, since at lower redshift more of the
gas will have condensed along the filaments (Dav{\'e} \etal 1999), the
two samples discussed here probably do not cover a wide enough range
in redshift to see this. Our $b$ values go up to about 150 \kms ,
unusually large for Ly$\alpha$ forest absorption lines, although a few
of the Bowen \etal (2002) sample do reach similar values. If the lines are
indeed composed of only a single component, and assuming that the
linewidth is purely due to the gas kinetic temperature (in which case
$b = \sqrt{2kT/m_p}$ with $k$ being the Boltzmann constant and $m_p$ the proton mass) , then this
would represent absorption from hot gas at a temperature of
$T=60.6b^2$, in our cases $T=0.4$ to $1.4\times10^6$K.

Based on these relationships it thus appears that the cosmic web is
the most likely explanation for the origin of our Ly$\alpha$ 
absorption-line detections. Thus the positive Ly$\alpha$
detections, despite appearing in proximity to the target galaxies, are
not directly related to these galactic disks, but only indirectly
related because of the way that these galaxies are positionned in the
large scale structure of the cosmic web.

\section{Summary and Conclusions}

Sightlines of QSOs at impact parameters 55 to 387 $h^{-1}_{75}$~kpc from nine nearby
galaxies (at systemic velocities from 566 to 5538 \kms) were probed,
using GHRS and STIS. In five cases a Ly$\alpha$ absorption line was
successfully detected at a velocity coincident with the galaxy's.
Some of these Ly$\alpha$ lines are the lowest-redshifts Ly$\alpha$
lines ever detected, and in fact most of them are sitting on the wing
of the Milky Way Ly$\alpha$ damped line (hence introducing
uncertainties in the derived equivalent widths).  The positive
detections occur for QSO-galaxy pairs with separations from 55 to 169
$h^{-1}_{75}$~kpc, while non-detections are for pairs with separations from 162 to
387 $h^{-1}_{75}$~kpc. Although at first sight it appears that the Ly$\alpha$ lines
might be genuinely arising in the extended disks of the target
galaxies, this is not what transpires from inspecting the detailed
kinematics of these galaxies.  HI velocity fields were obtained at the
VLA for our galaxies (and in one case was available from the
literature), to derive their rotation curves. When comparing the inner
rotation curves of the galaxies with the velocity at large radius
provided by the Ly$\alpha$ line it appears that it is very difficult
to explain this Ly$\alpha$ velocity as part of the extended gaseous
rotating disk.  In most cases one would need to invoke large warps in
the extreme outer gas disks to reconcile the observed velocities with the
predicted ones. Worse, in some cases the Ly$\alpha$ line velocity
indicates in fact counter-rotation with respect to the inner disk
rotation.

In light of these results it appears that there is no detectable gas,
down to levels of about 10$^{13}$ cm$^{-2}$, corotating in an extended gaseous 
disk at radii $> 35$ $\alpha ^{-1}$.  
The cosmic web is the most likely origin for the detected Ly$\alpha$
lines.  The observed equivalent widths of the lines are consistent
with this picture.  Indeed the equivalent widths are correlated with
the local volume density of galaxies around the sightline. This makes
sense if the Ly$\alpha$ lines arise from the cosmic web, which is
denser in regions of higher volume density of galaxies (since the
galaxies have formed by condensating in the denser parts of the
network). This correlation would be difficult to explain if the
Ly$\alpha$ lines arised from halos or extended disks of the most
nearby galaxy. One could argue that the equivalent widths should be
larger for sighlines close to larger galaxies, which tend to be found
in regions of higher galaxy volume density, which could then explain
the trend observed.  However if this were the case then one would
observe presumably an even stronger correlation between the equivalent
widths and the luminosity of the nearby galaxy, and although this was
reported to be the case by Chen \etal (1998) it is now clearly
contested by the results of Bowen \etal (2002), as well as in our sample 
for which no correlation is found between equivalent widths and luminosities.

Note that these results do not necessarily imply that the dark matter haloes of these galaxies do not
exist at the radii probed by our QSO sightlines. It simply means that there is no 
longer some detectable gas in rotation in an extended gas disk at these radii,
and that the Ly$\alpha$ lines detected are due to foreground or background cosmic
web gas surrounding the galaxy. In fact weak lensing studies seem to indicate
that indeed the dark matter halo extend very far out, typically $>$ 300 
$h^{-1}_{75}$kpc for
a $L_*$ galaxy (eg. the weak-lensing studies such as Smith \etal 2001). But there is no gas to trace its dynamics
out there, at least down to our limiting column densities. 
HI as traced by background QSOs is therefore not a useful tracer of galactic 
potential in the far outer regions of a $\sim $200 kpc halo. Only weak lensing,
and perhaps H$\alpha$ in recombination from the extended ionized disk 
(Bland-Hawthorn \etal 1997) are hopeful tracers of dynamics at such radii.
These QSOs 
turn out however to be extremely useful to probe the phases of the cosmic web gas.
Many more sightlines will be required though before one can attempt to
understand the cosmic gas distribution and density, as well as its evolution
with redshift. The planned HST COS spectrograph would have allowed to probe the sightlines
of fainter QSOs which would greatly increase the available sample, and hopefully the decision regarding HST can be reversed in the future. Eventually the
next generation of sensitive giant radio telescopes as the CLAR (C\^ot\'e \etal 2002)
and eventually
the SKA, will be able to survey
efficiently large areas of the sky down to colum densities as deep as 10$^{16}$
cm$^{-2}$, and hence trace directly in emission the
structure of such low column density gas in the cosmic web.

\acknowledgments

We wish to thank an anonymous referee for thorough comments that helped
improve this manuscript.
Ray Lucas and especially Jennifer Wilson are thanked
for data reduction support during visits at STScI.
SC thanks Jacqueline Bergeron for useful insights on QSO absorption lines.
 The National Radio Astronomy Observatory is a facility of the National Science Foundation operated under cooperative agreement by Associated Universities, Inc. 
This reasearch has made use of the Canadian Astronomy Data Center, which is operated  
by the National Resarch Council of Canada's Herzberg Institute of Astrophysics.
Extensive use of the NASA/IPAC Extragalactic Database (NED), operated by the Jet
Propulsion Laboratory, California Institute of Technology, under contract with 
NASA, is acknowledged.
The HIPASS data used here were obtained at the Parkes Telescope, part of the
Australia Telescope funded by Australia for operation as a National Facility
managed by CSIRO.
Support for this work was provided by NASA through grant numbers
GO-6665.02-95A and GO-07295.02-96A from the Space Telescope Science
Institute, which is operated by AURA, Inc. under NASA contract
NAS5-26555.

\clearpage

\begin{figure}
\centering
\includegraphics[width=8cm]{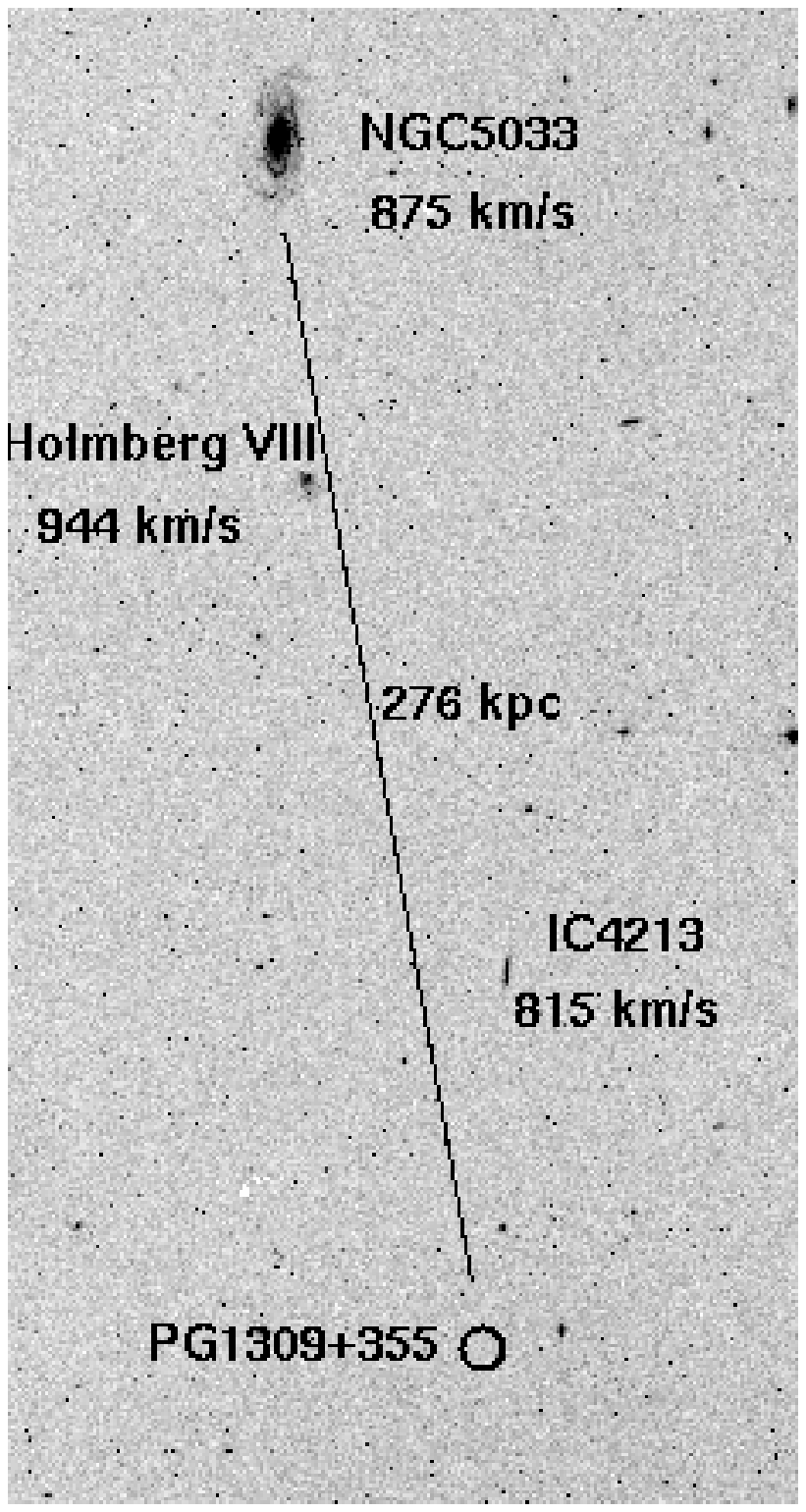}
\caption{Digitized Sky Survey images (North up, East left) for our 5 galaxy-QSO
pairs for which a Ly$\alpha$ line was detected. The systemic velocity of the 
galaxy is indicated, as well as the projected distance to the QSO sightline.
In the case of PG1309+355 there are also several dwarf galaxies close to the QSO
sightline.
 \label{fig1} }
\end{figure}
\clearpage

\clearpage


\begin{figure}
\plotone{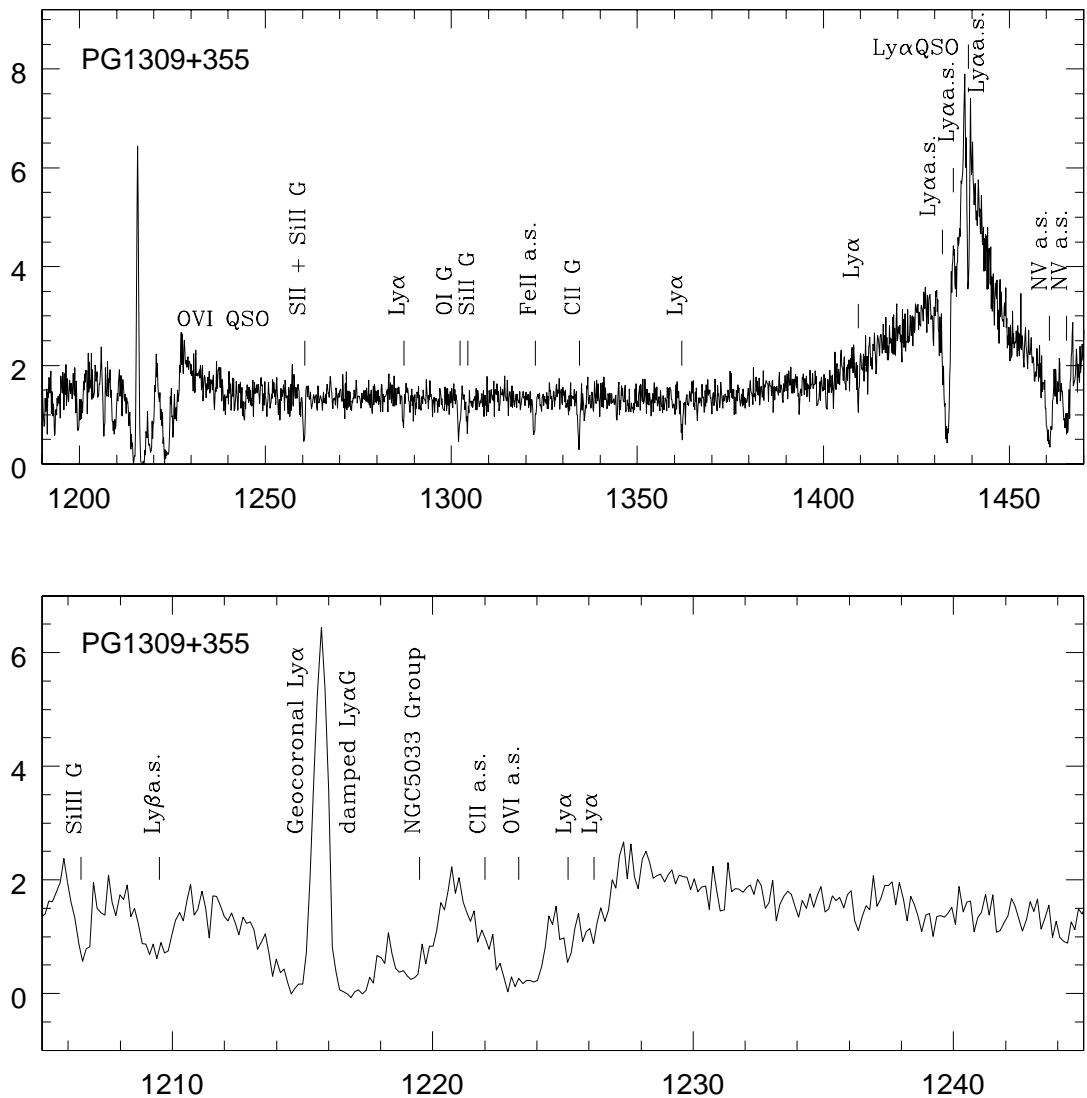}
\caption{a) GHRS G160L spectrum of PG1309+355 and b) STIS G140M spectra for the other
QSOs, with the
detected Ly$\alpha $ absorption lines associated with our target galaxies. 
Lines indicated with "G" are galactic, those with "a.s." are from the 
associated absorption systems found around PG1309+355. Other  Ly$\alpha $ 
lines are found at other various redshifts as well. c) Spectra of the QSOs
with no detection of a Ly$\alpha $ absorption line associated with
the target galaxy: GHRS G160L for PG0923+201 and STIS G140M for the 
others \label{fig2} } 
\end{figure}
\clearpage
\begin{figure}
\plotone{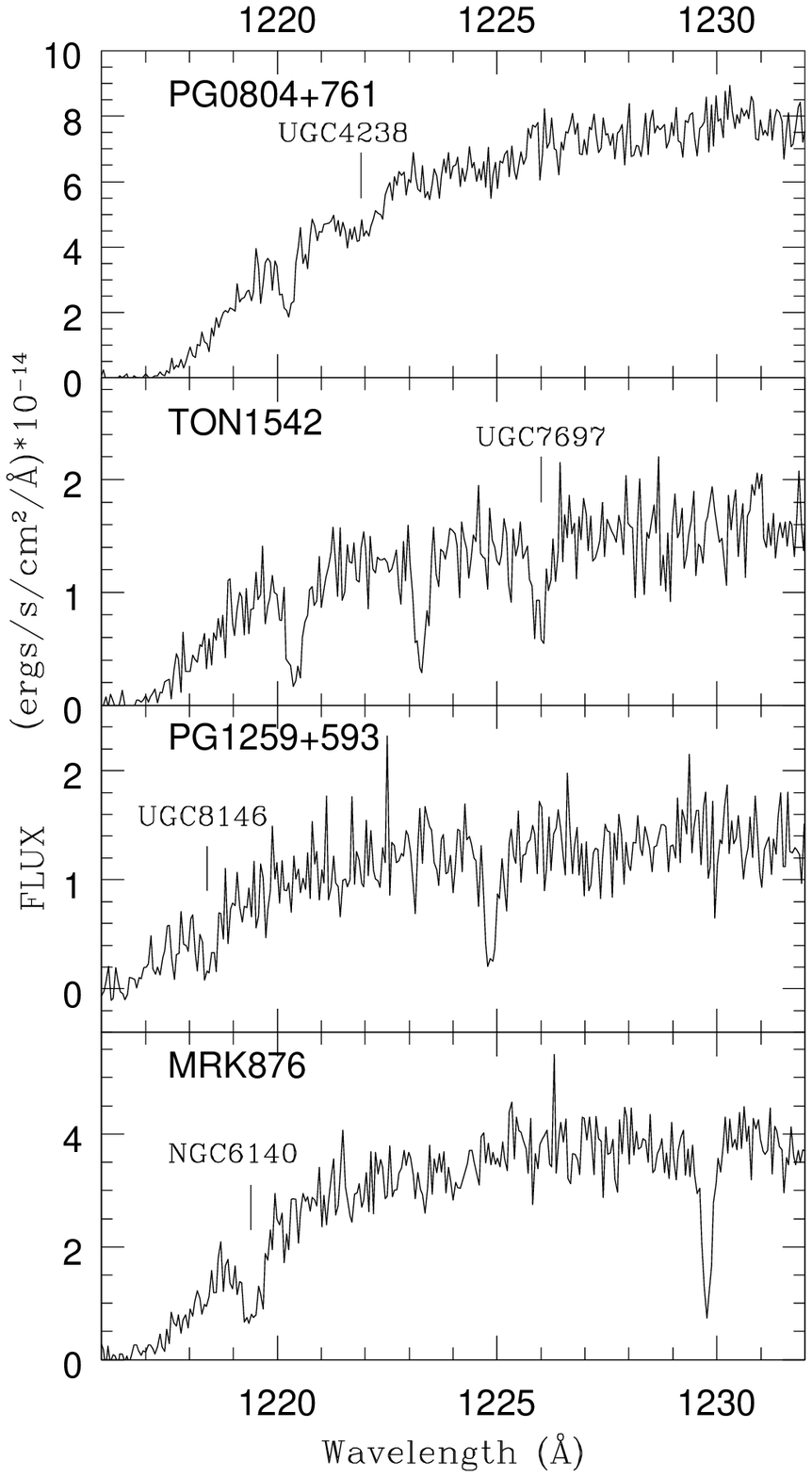}
\end{figure}
\clearpage
\begin{figure}
\plotone{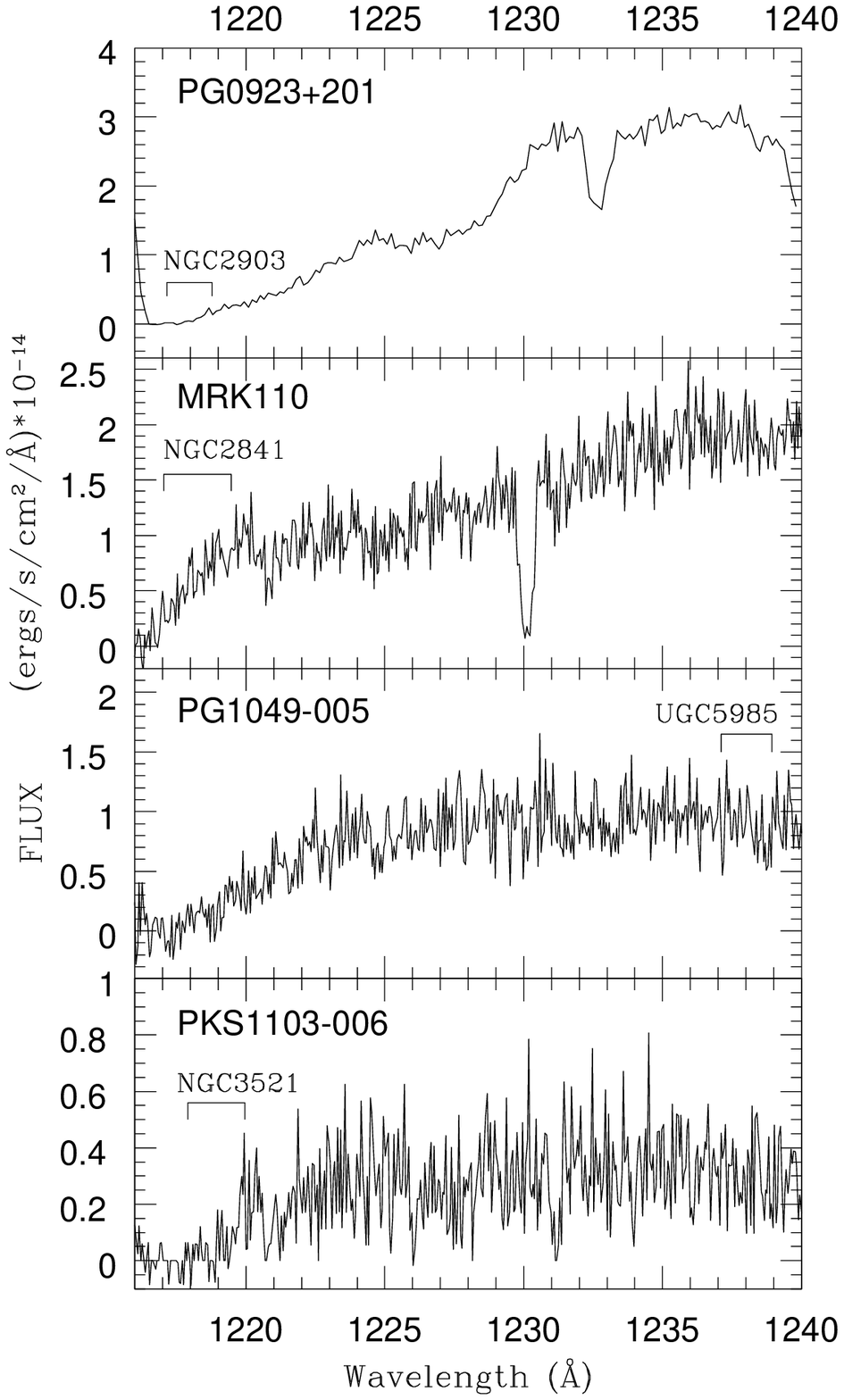}
\end{figure}

\clearpage 
\begin{figure}
\epsscale{0.7}
\plottwo{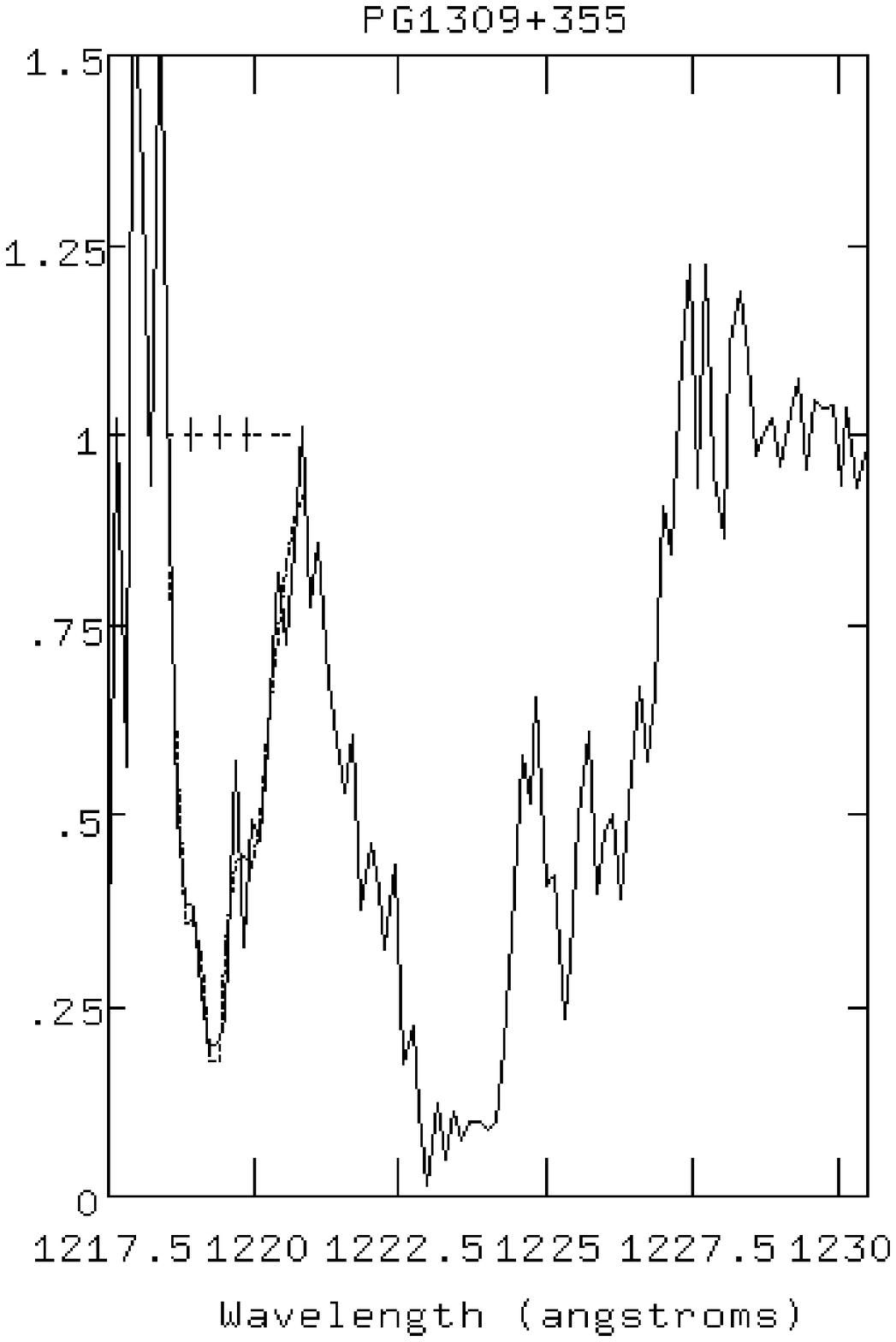}{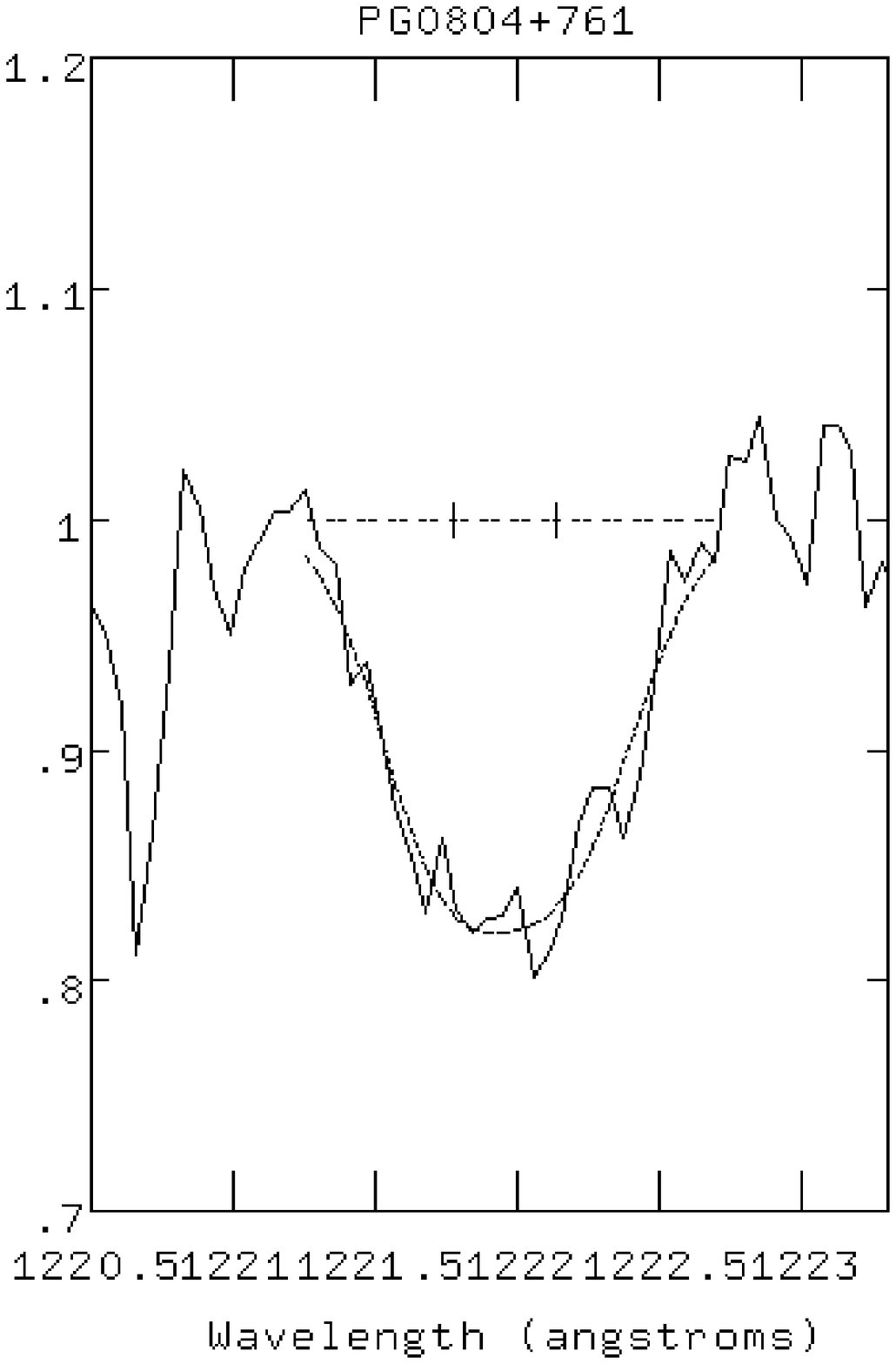}
\caption{Normalized spectra with profile fits on the detected absorption lines
(the subcomponents for PG1309+355 and PG0804+761 are shown as vertical ticks)
 \label{fig3} }
\end{figure}

\clearpage
\begin{figure}
\epsscale{0.6}
\plottwo{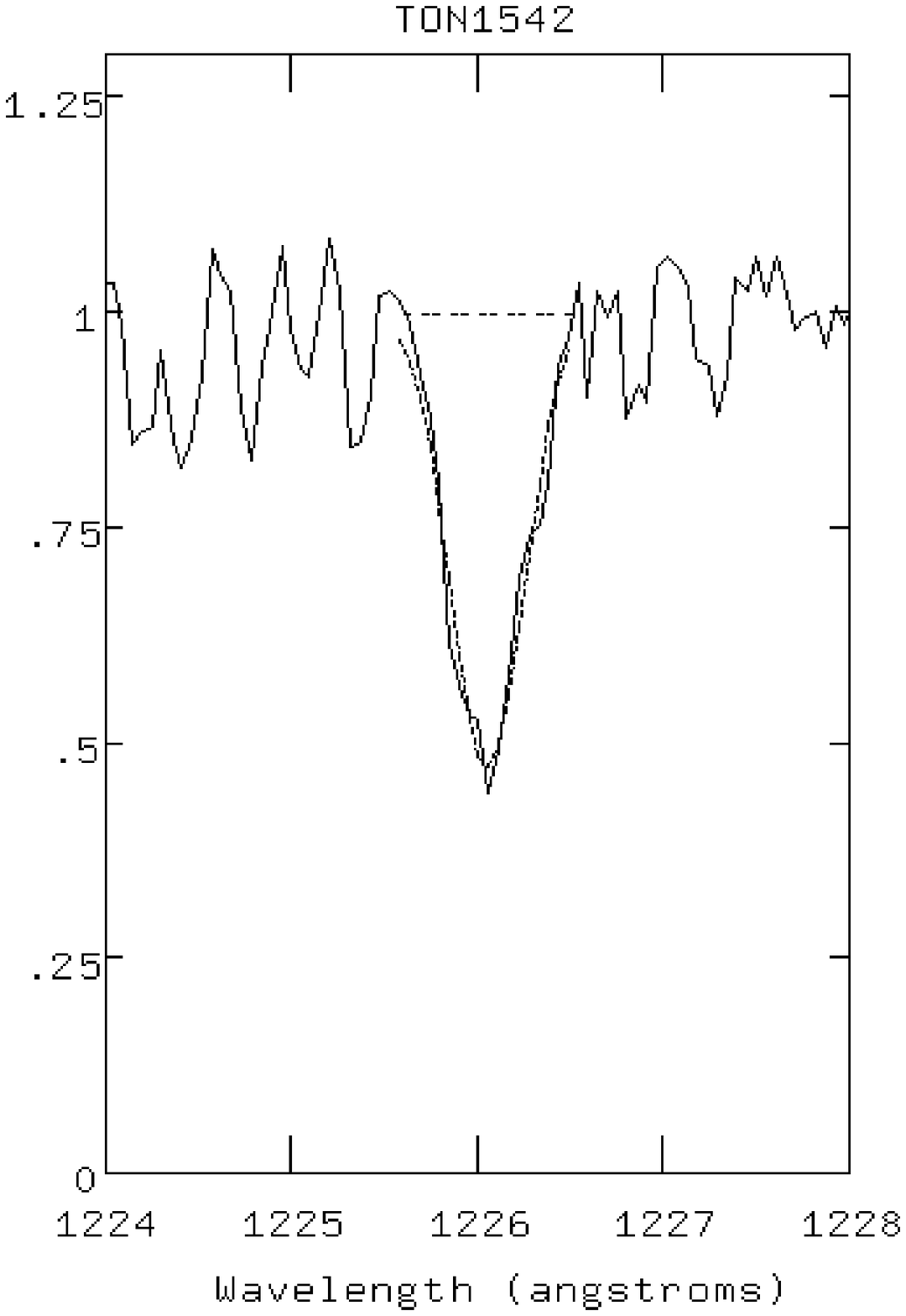}{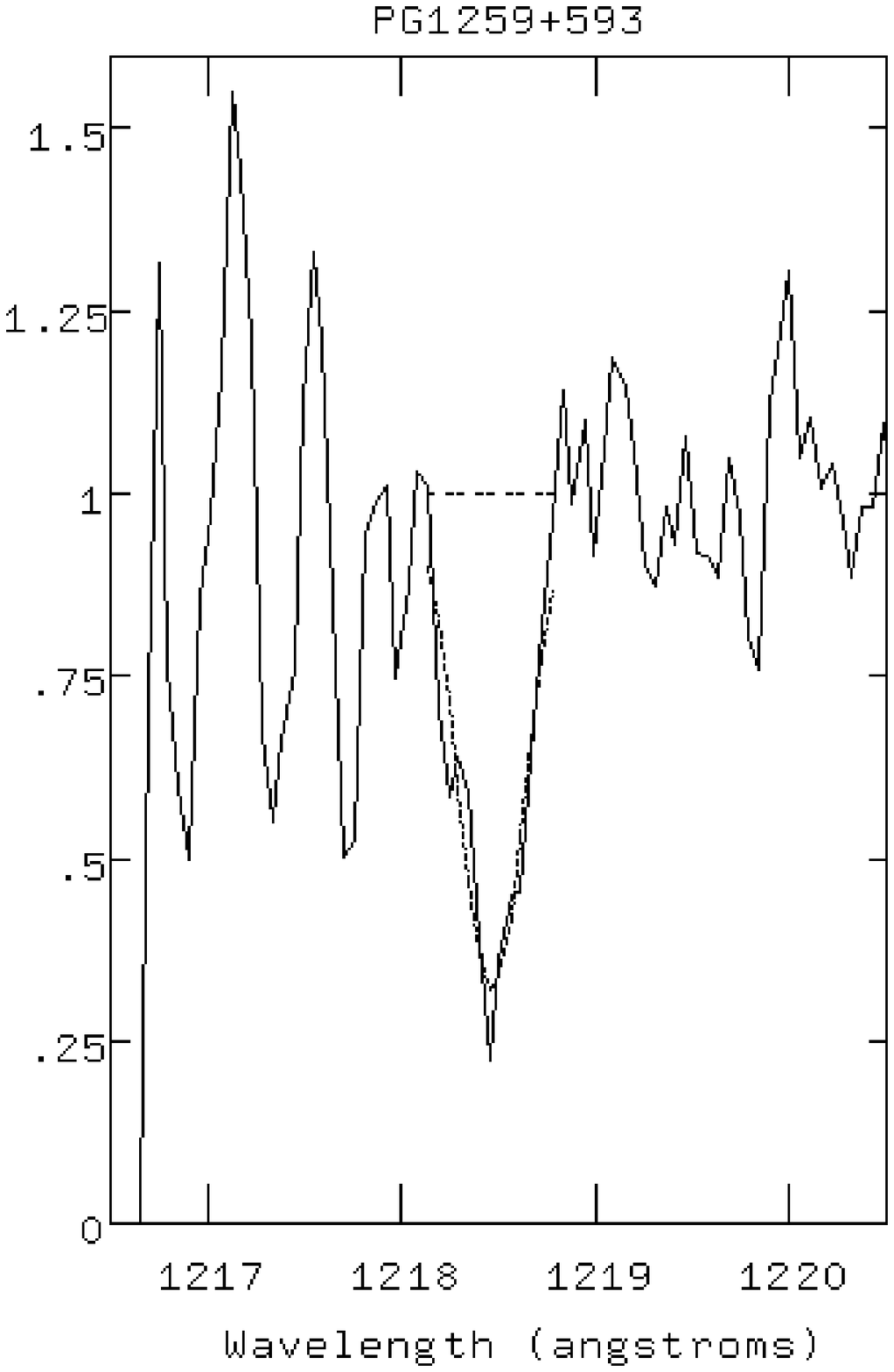}
\plotone{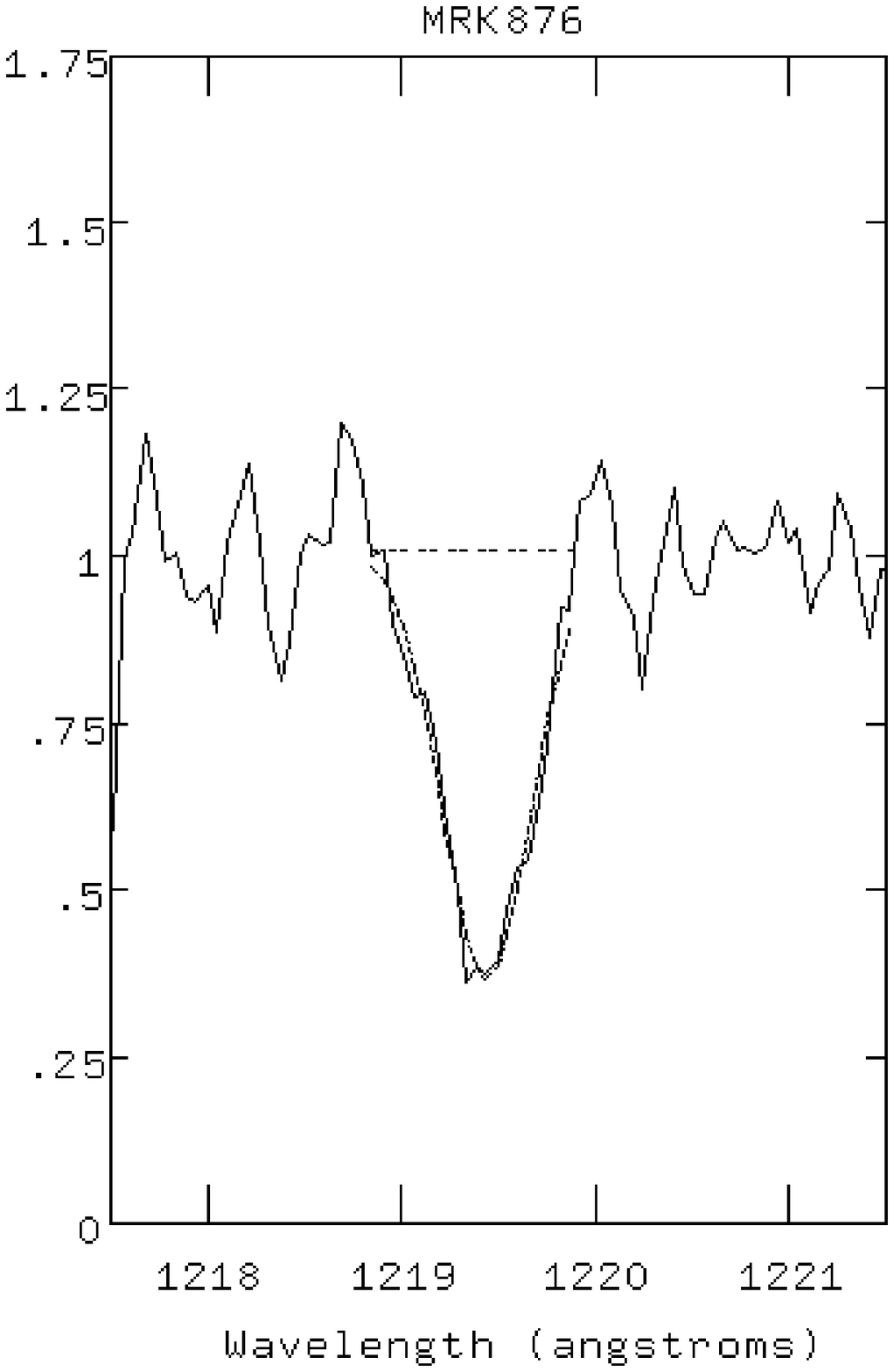}
\end{figure}

\clearpage 

\begin{figure}
\centering
\includegraphics[width=12cm]{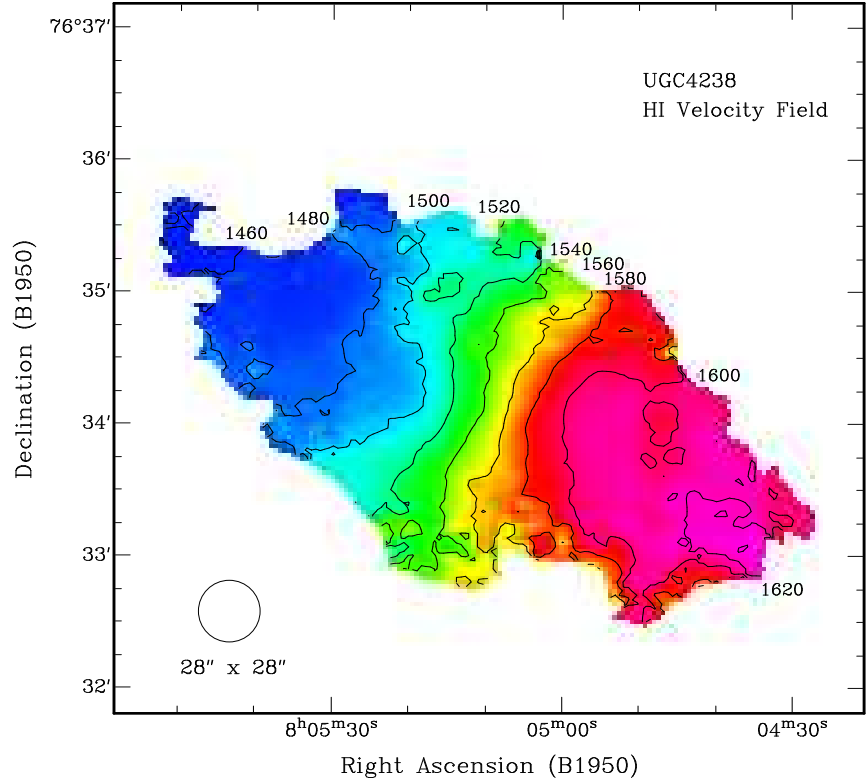}
\caption{VLA HI velocity field and HI column density distribution over a DSS image, for UGC4238. The
isovelocity contours are at interval of 20 \kms, and for the column density map
the contour levels are at 3.2, 9.5, 16.0, 22.0 and 28.4$\times 10^{20}$ atoms cm${-2}$.
Same for UGC7238, with contour levels at 0.4, 1.3, 2.1, 3.0 and 3.8$\times 10^{20}$
atoms cm${-2}$. Same for NGC6140, with contour levels at 0.89, 2.7,
4.4, 6.2 and 8.0$\times 10^{20}$atoms cm${-2}$. The beam sizes are indicated in the lower corners of the figures. \label{fig4} } 

\end{figure}
\clearpage
\begin{figure}
\centering
\includegraphics[width=18cm]{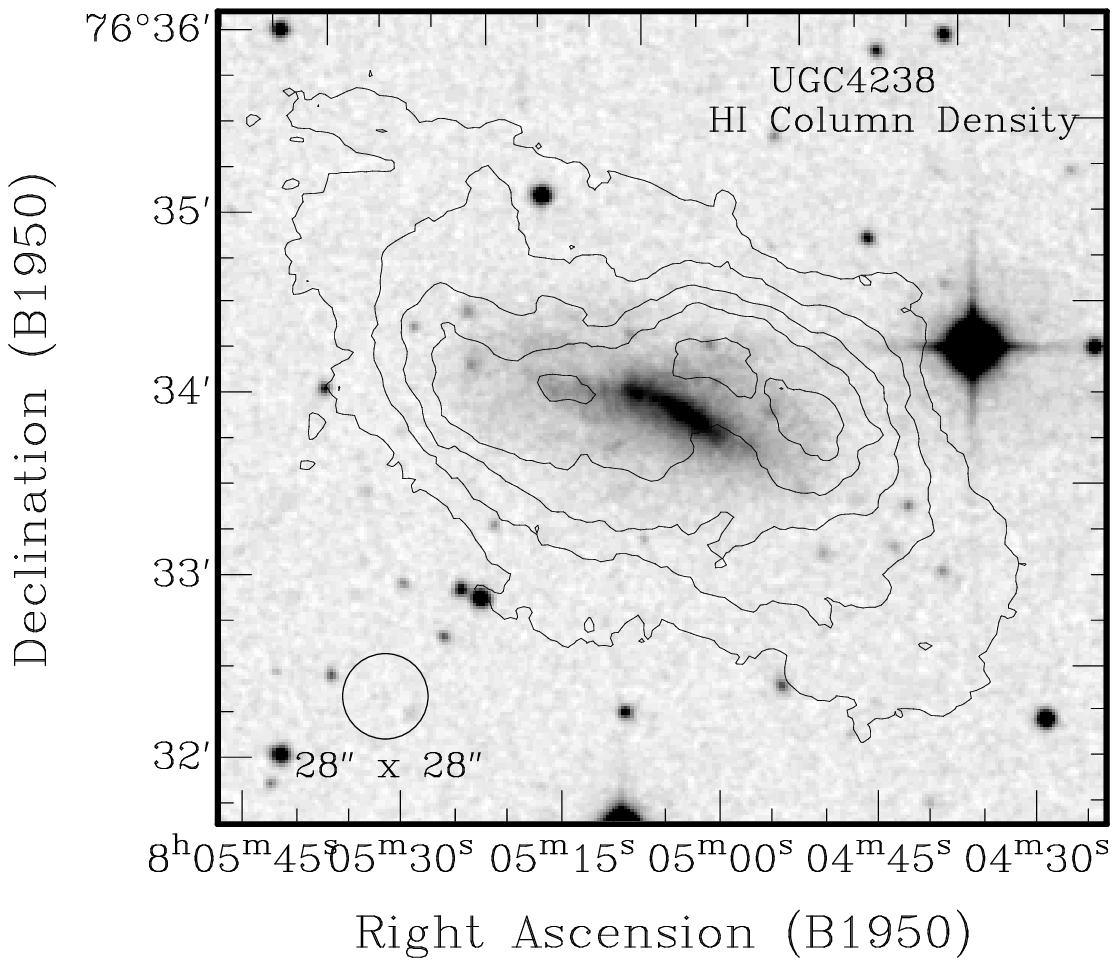}
\end{figure}
\clearpage
\begin{figure}
\centering
\includegraphics[width=12cm]{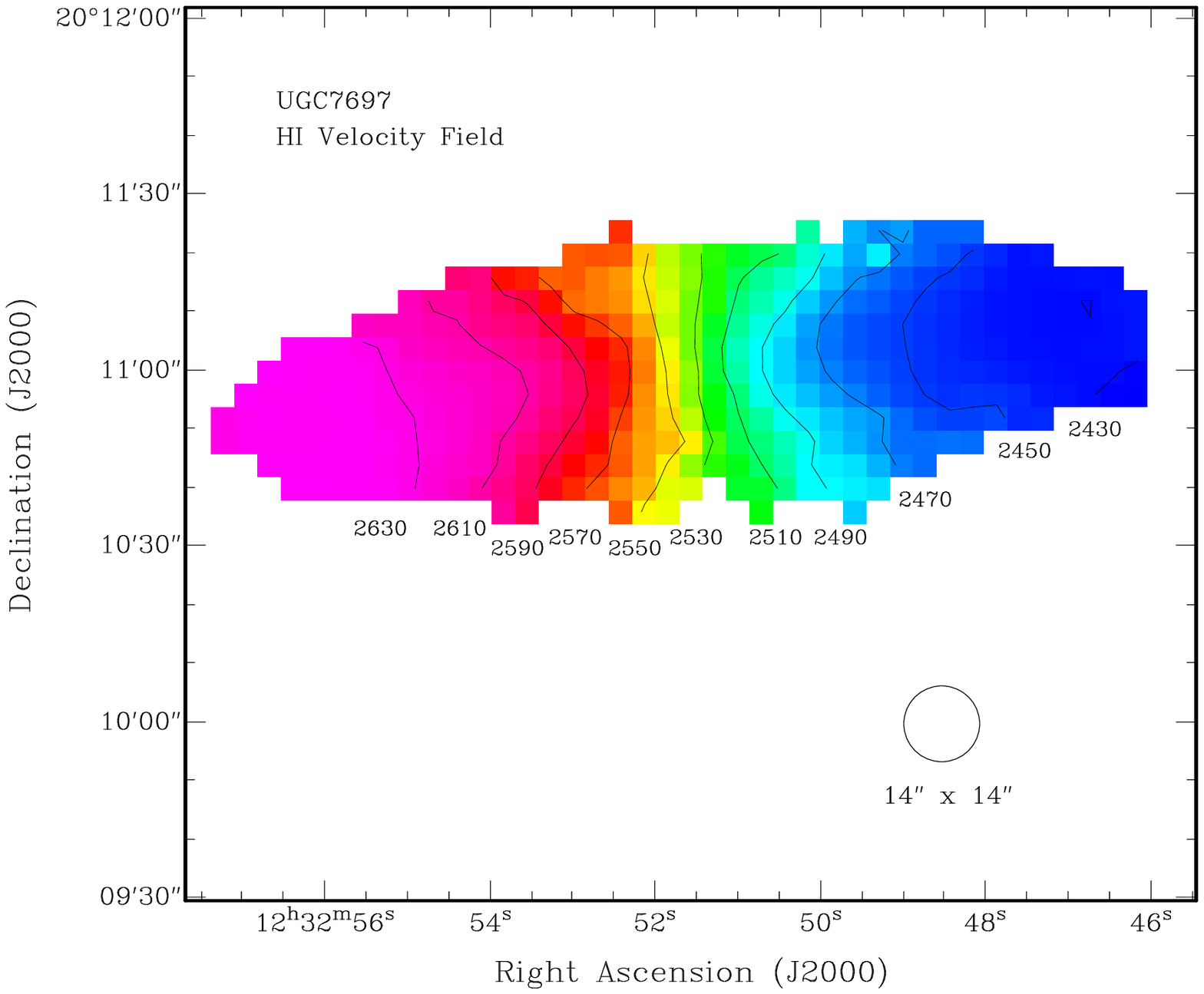}
\end{figure}
\clearpage
\begin{figure}
\centering
\includegraphics[width=18cm]{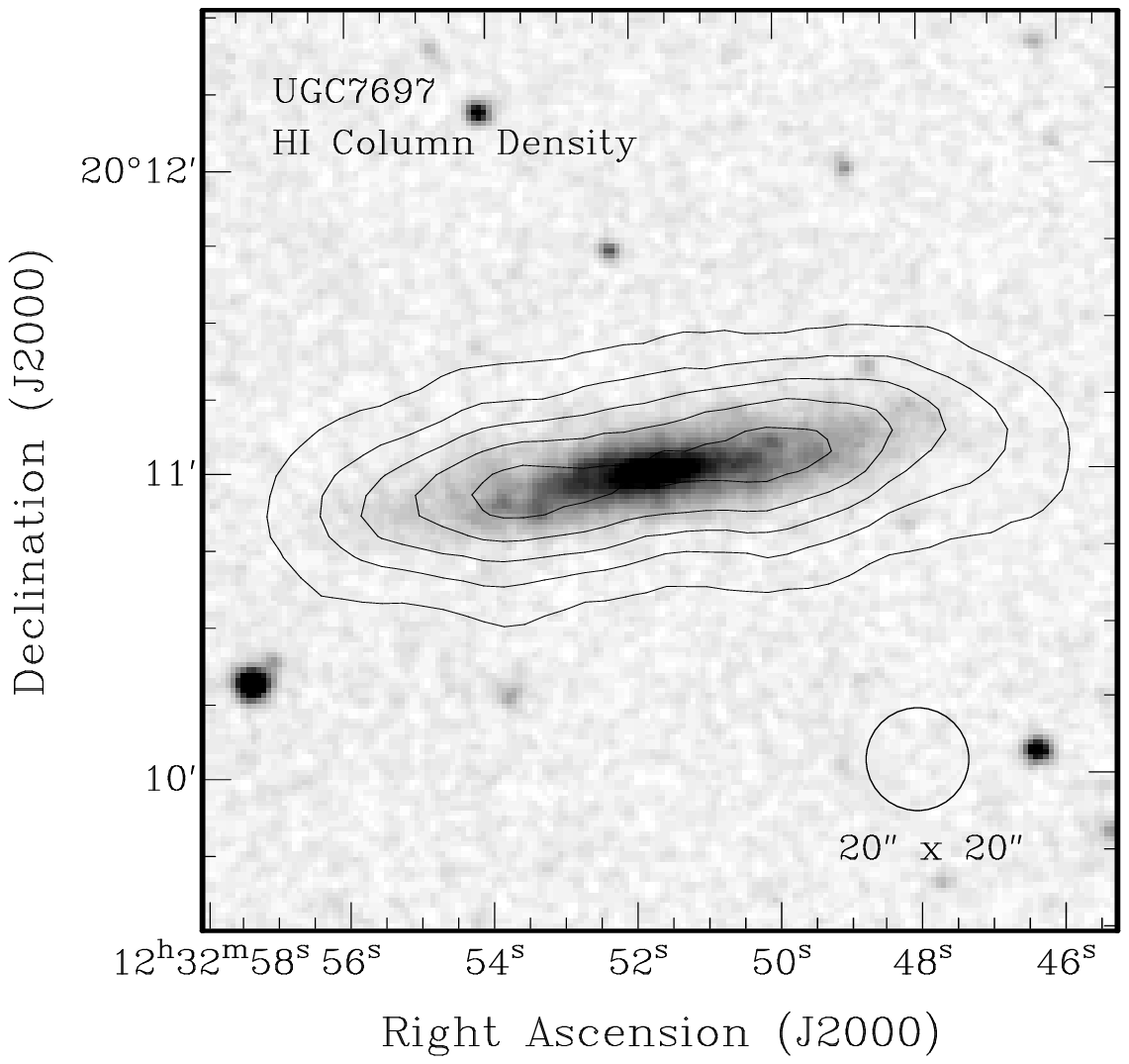}
\end{figure}
\clearpage
\begin{figure}
\centering
\includegraphics[width=12cm]{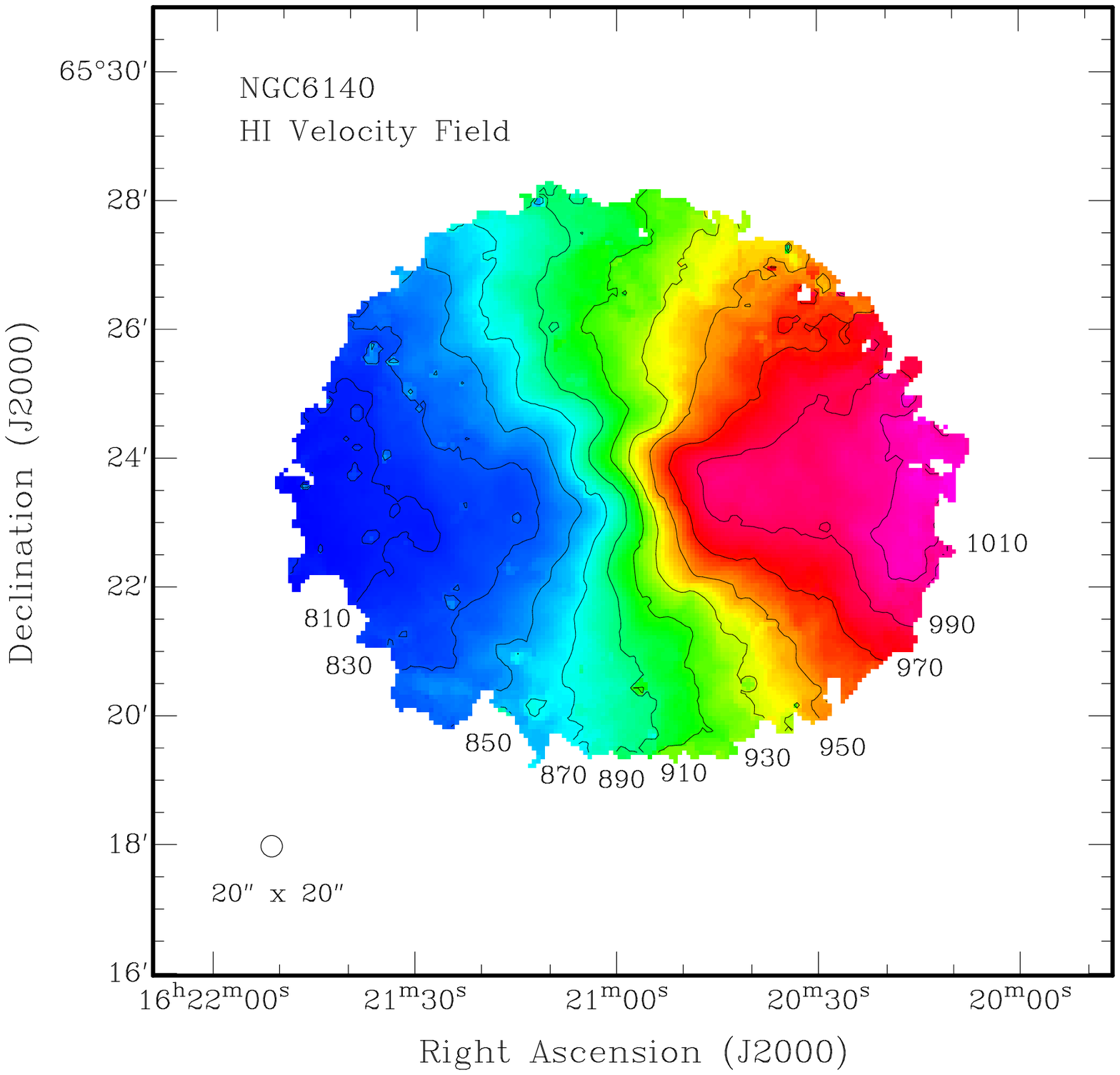}
\end{figure}
\clearpage
\begin{figure}
\centering
\includegraphics[width=15cm]{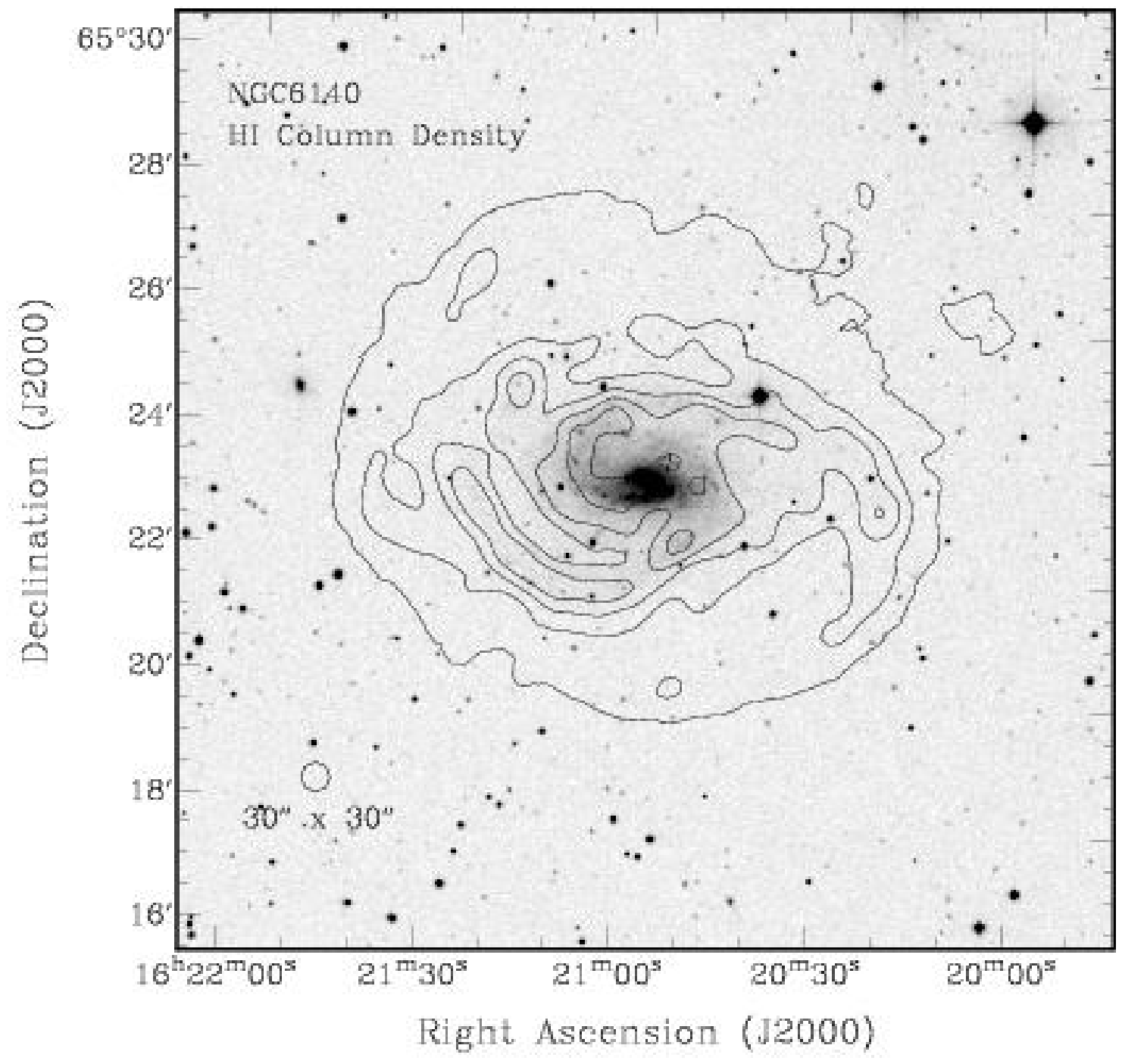}
\end{figure}
\clearpage

\clearpage

\begin{figure}
\epsscale{1.0}
\plotone{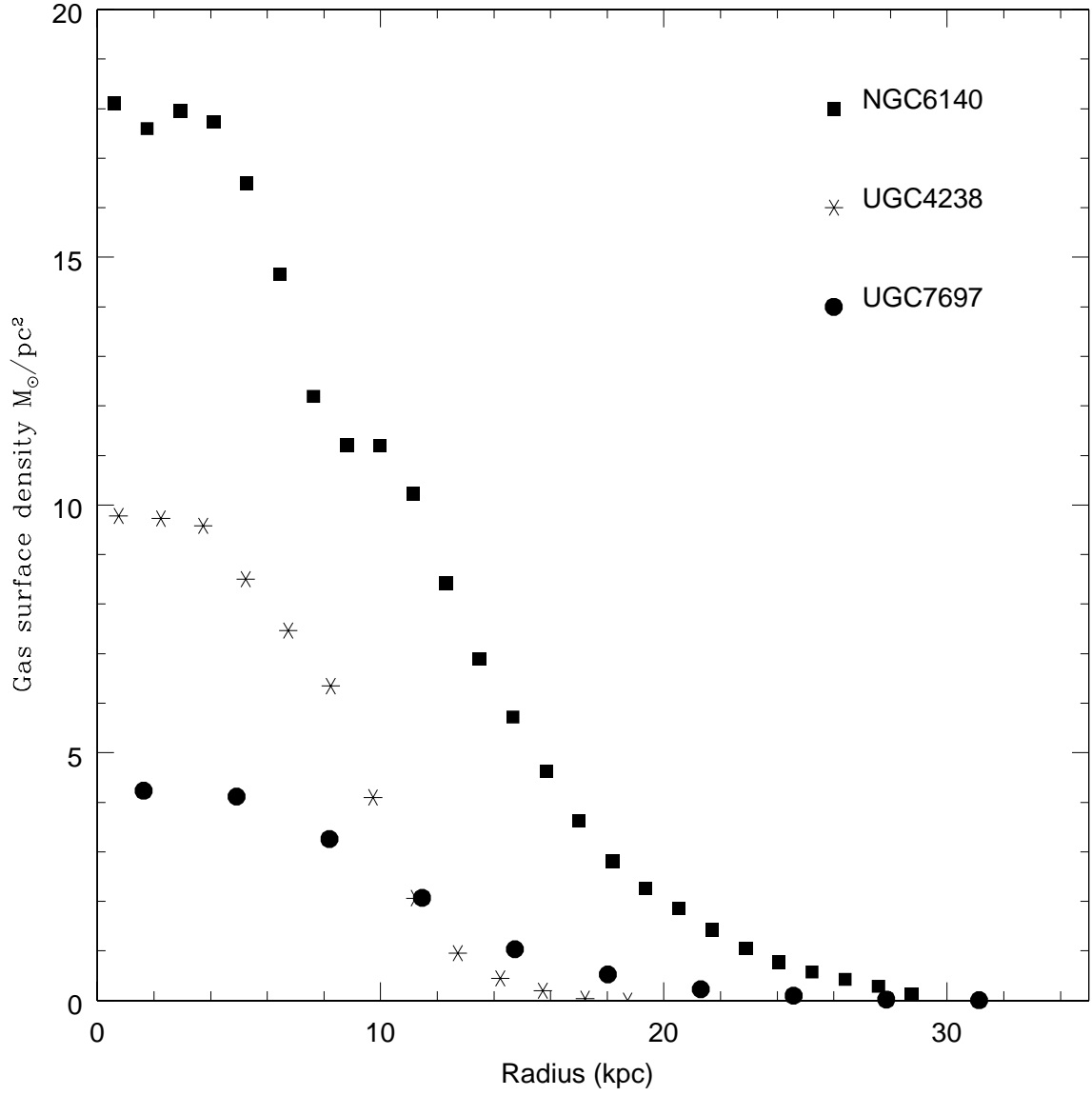}
\caption{Gas radial surface density distributions for UGC4238, UGC7697 and 
NGC6140. \label{fig5} } 
\end{figure}

\clearpage 
\begin{figure}
\epsscale{1.0}
\plottwo{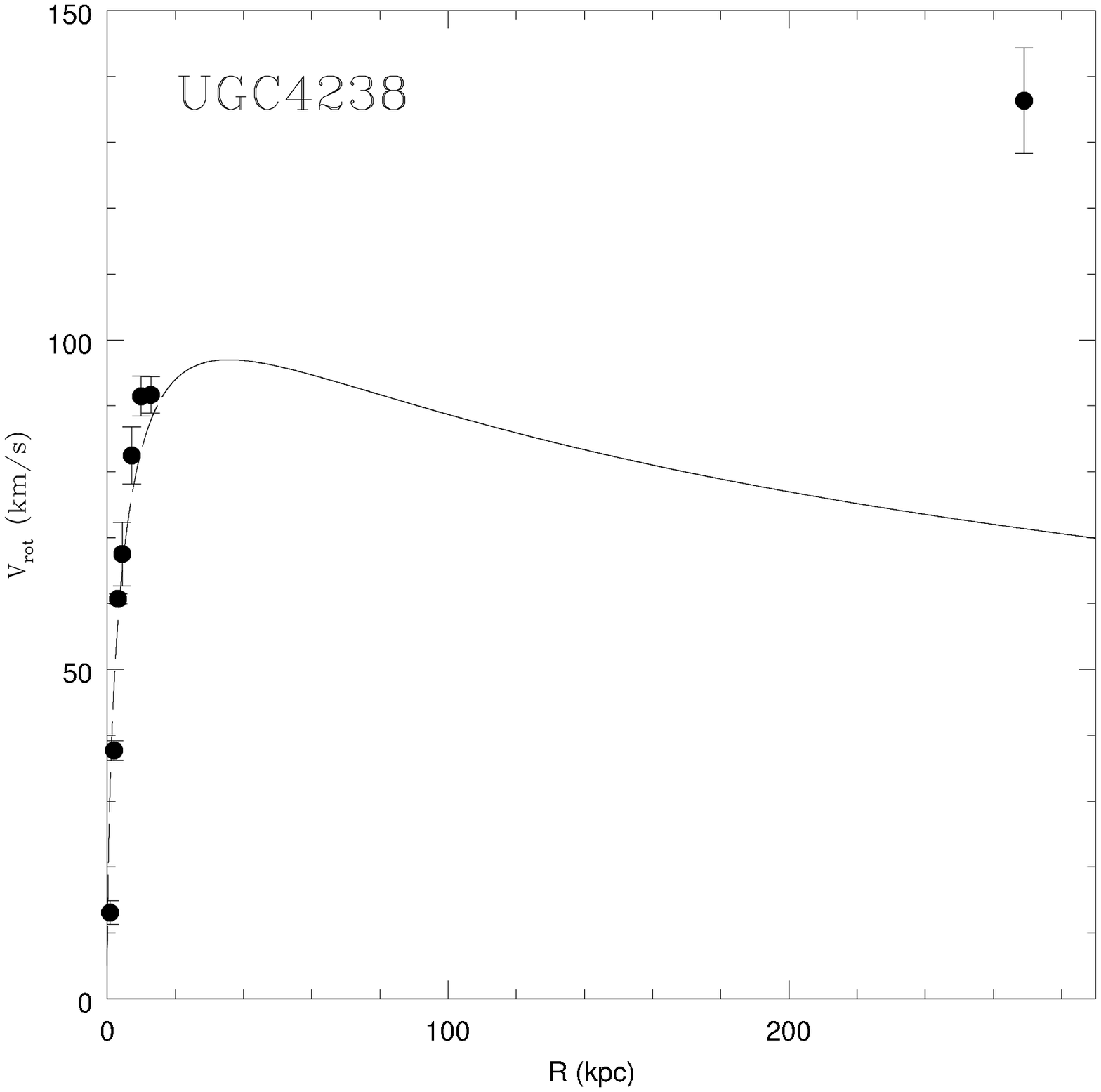}{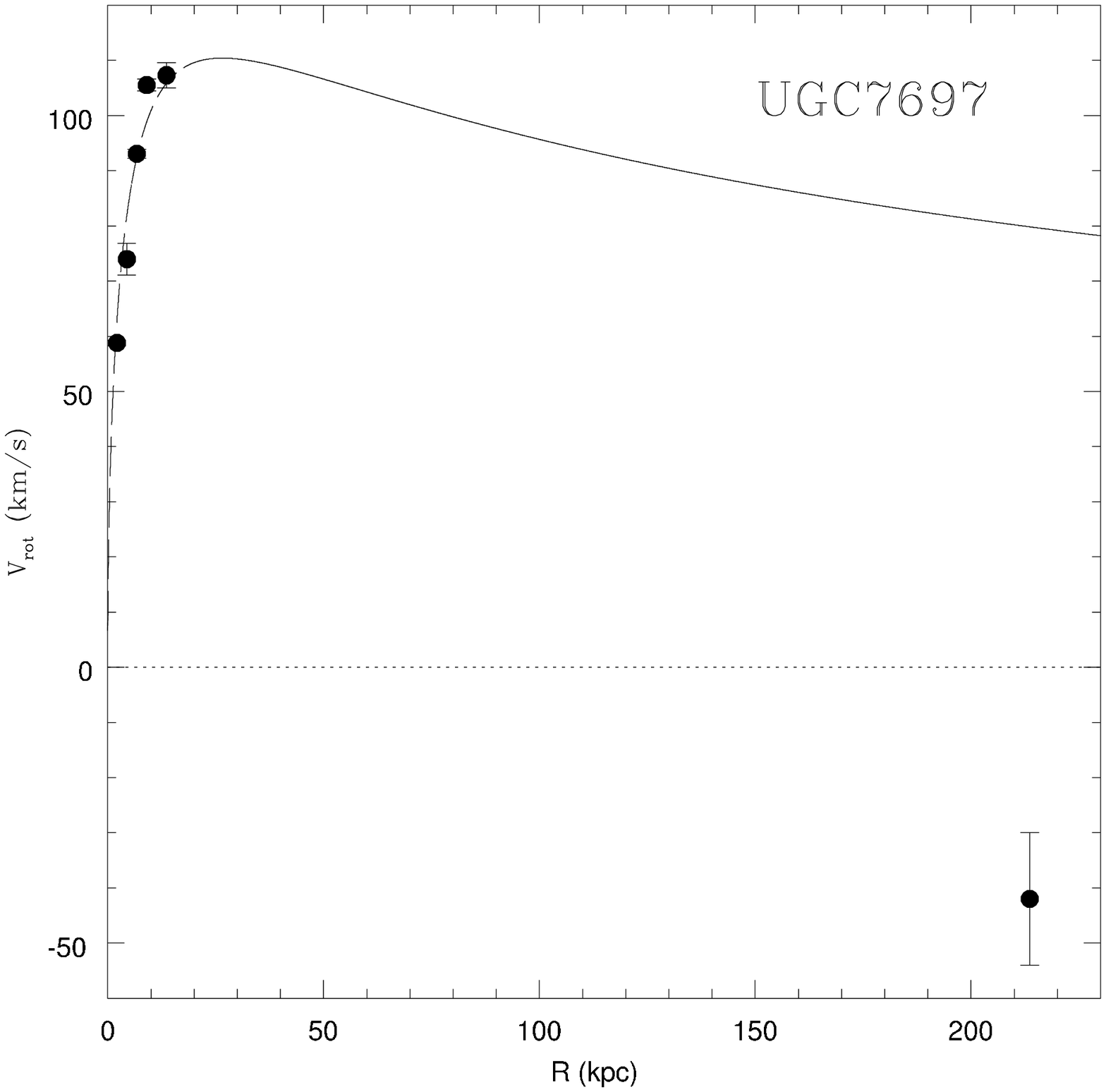}
\caption{Rotation curves, consisting of HI velocities in the inner parts and the
Ly$\alpha$ absorption line velocity at large radius. The line shows the velocities
expected by CDM. The observed Ly$\alpha$ velocities do not agree with those
expected for an extended corotating gaseous disk. \label{fig6} }
\end{figure}
\clearpage

\clearpage
\begin{figure}
\epsscale{1.0}
\plottwo{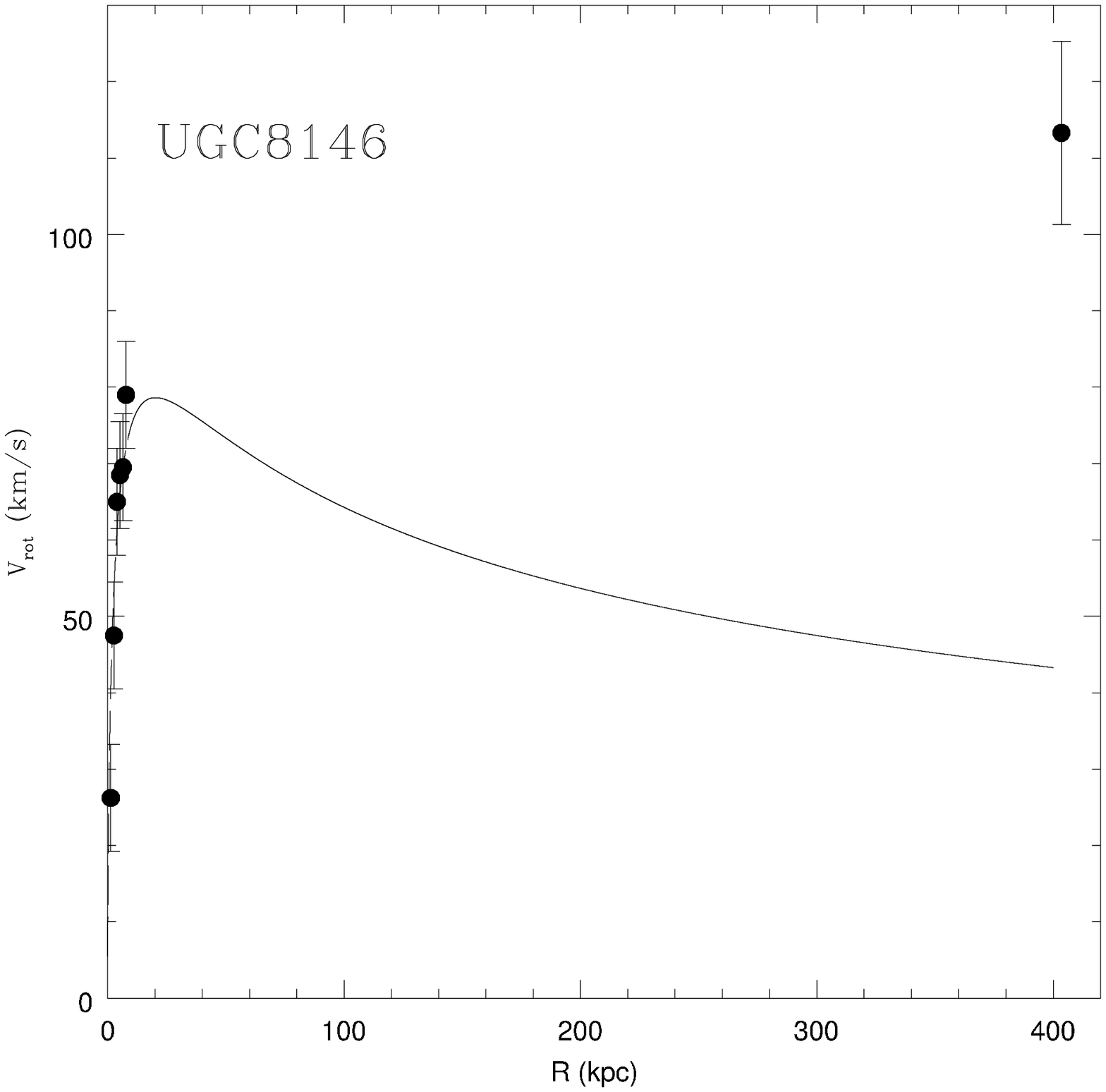}{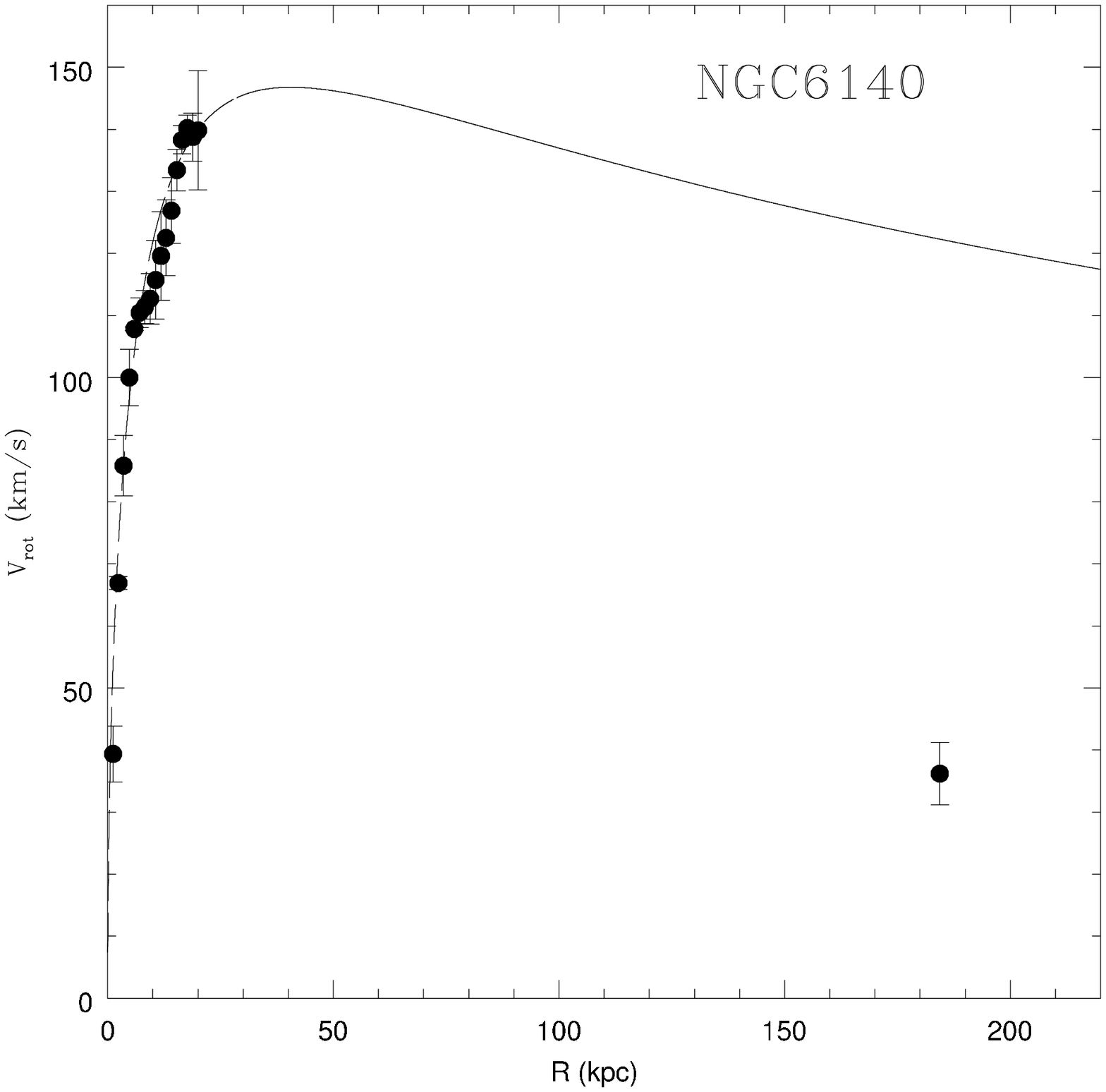}
\end{figure}
\clearpage

\begin{figure}
\epsscale{1.0}
\plottwo{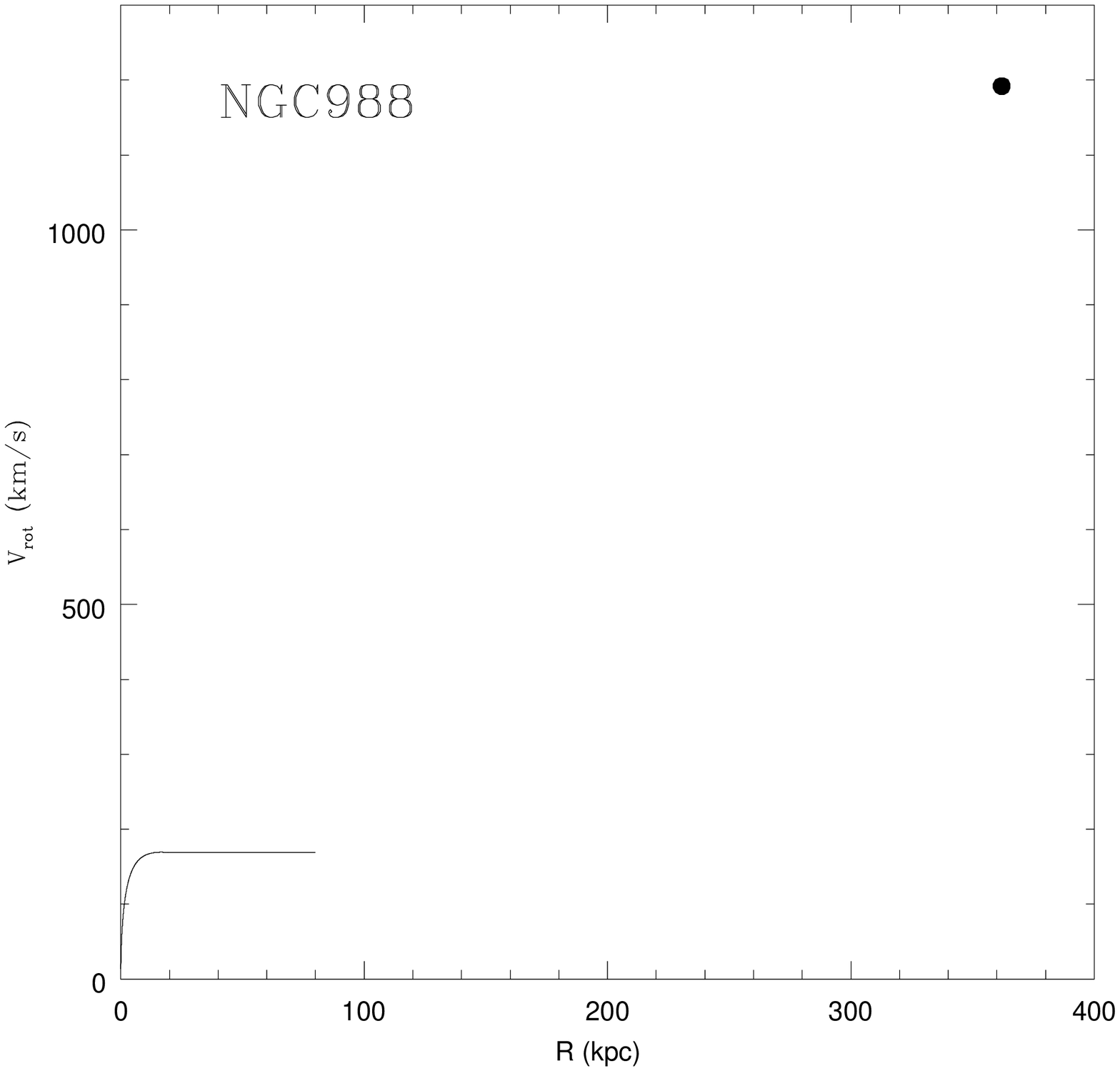}{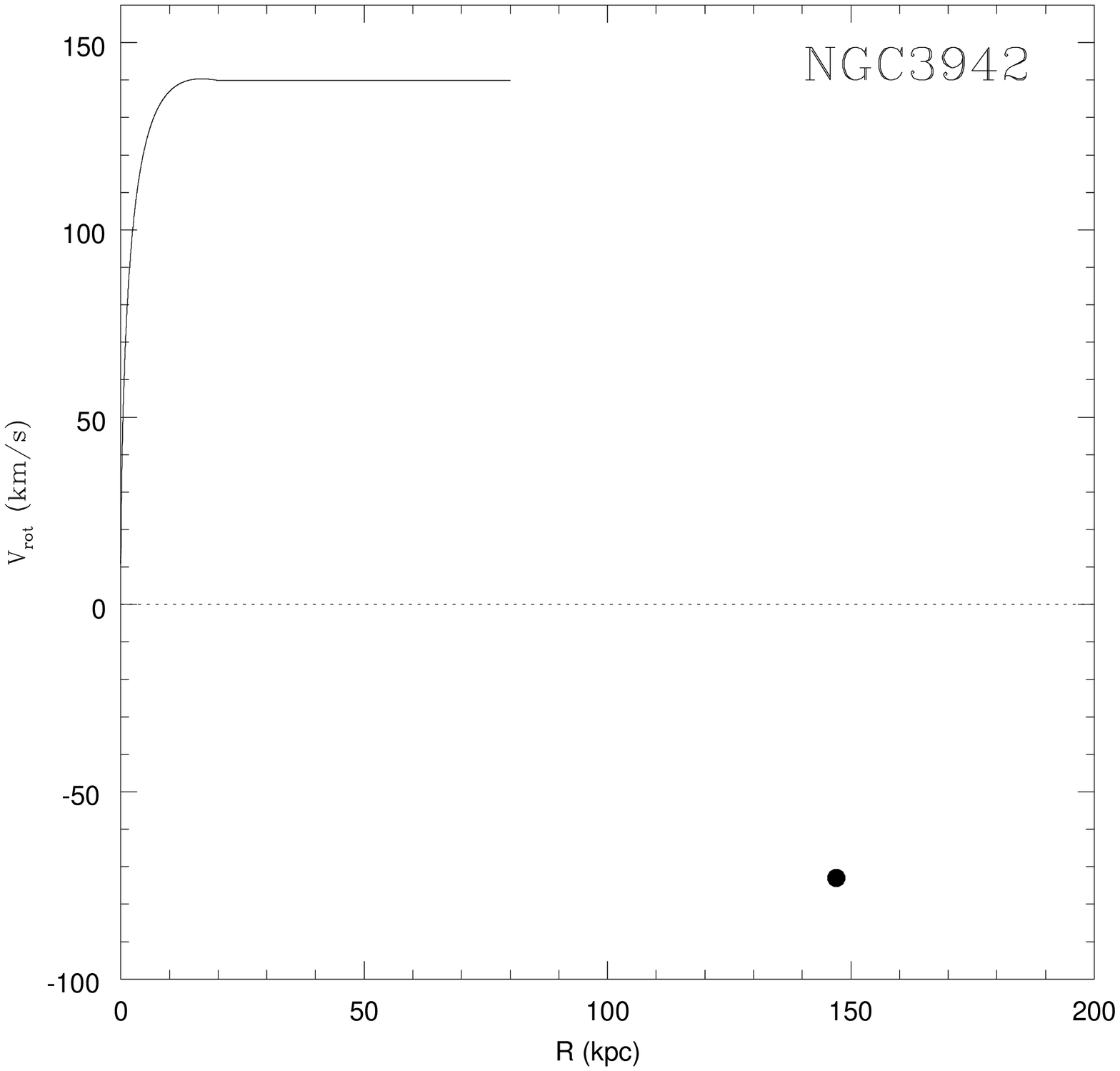}
\caption{Same as Figure 5, for two galaxies from the Bowen \etal (2002) sample.
 No rotation curves are available for these galaxies so  model ones based on the
'Universal Rotation Curve' of Persic \& Salucci (1996) are used. \label{fig7} }
\end{figure}

\clearpage
\begin{figure}
\plotone{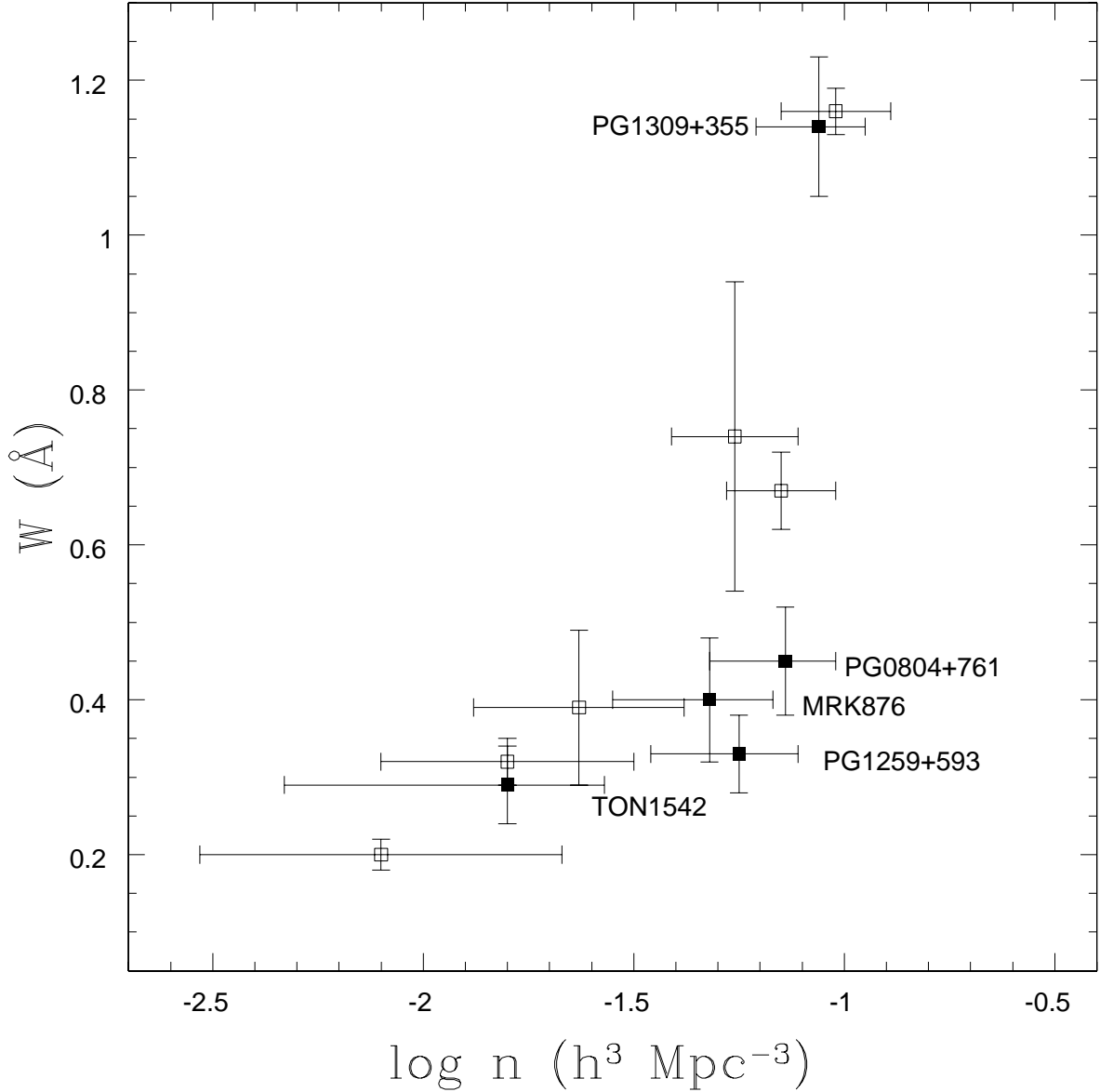}
\caption{Total equivalent width versus the volume density of galaxies brighter
than M$_B$=-17.5 in a cylinder of radius 2 $h^{-1}$ Mpc and length $\pm$500 \kms 
from the detected Ly$\alpha$ line. Filled squares are for our sightlines,
open ones for the sightlines of Bowen \etal (2002). \label{fig8} } 
\end{figure}

\clearpage

\begin{figure}
\plotone{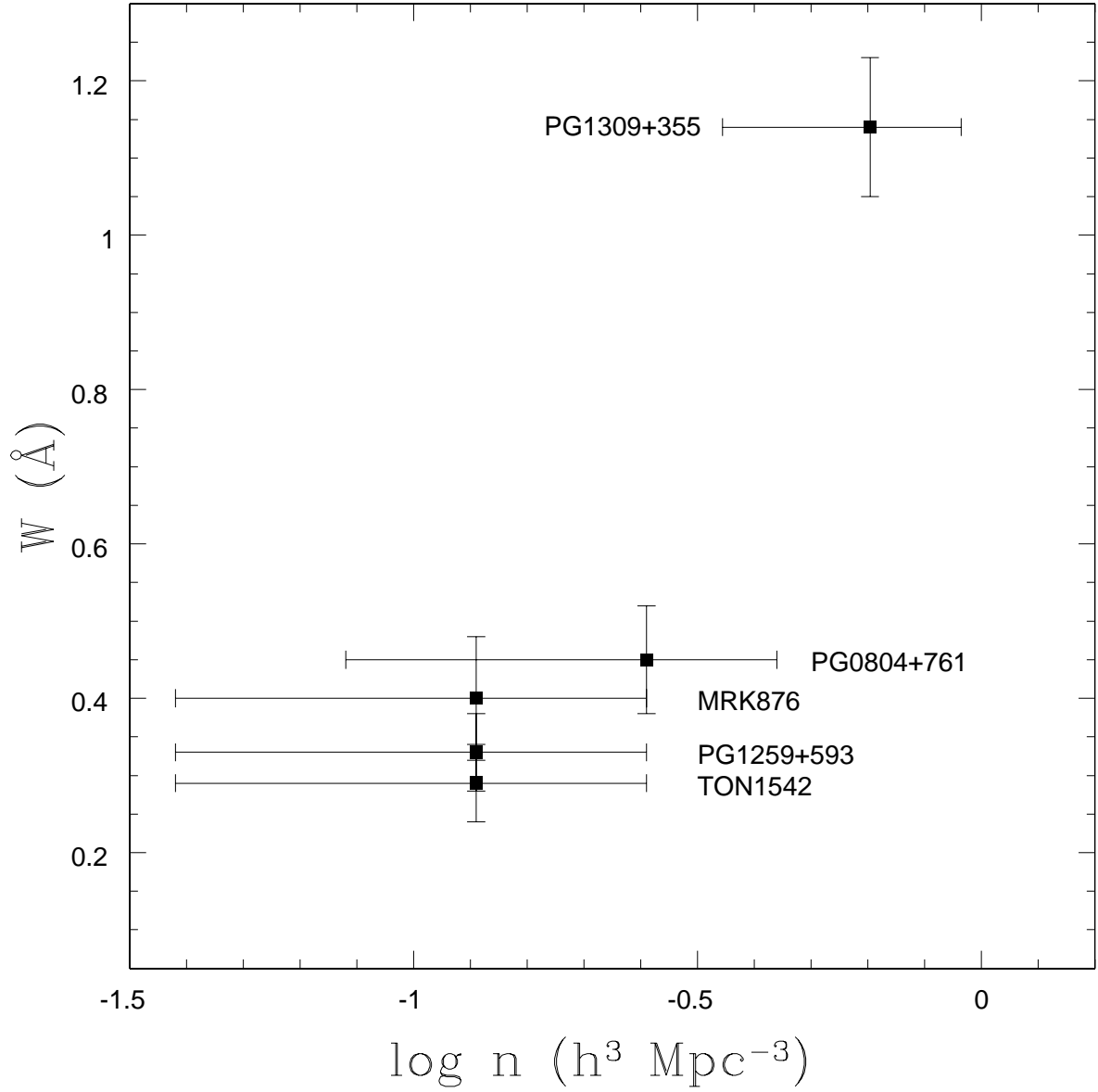}
\caption{Same as Figure 7, for galaxies brighter than M$_B$=-16.5
in a smaller cylinder of 0.5 $h^{-1}$ Mpc and length $\pm$500 \kms \label{fig9} } 
\end{figure}


\clearpage 

\begin{deluxetable}{cccccccc} 
\tablecolumns{8}
\tabletypesize{\small}
\tablecaption{Details of the HST Observations (PID=6665 \& 7295) \label{tab2}}  
\tablehead{
\colhead{Object} & \colhead{RA} & \colhead{Dec} & \colhead{Redshift} & 
\colhead{Instrument} & \colhead{Dataset} & \colhead{Date} & 
\colhead{Exposure time} \\
\colhead{}  & \colhead{(J2000)} & \colhead{(J2000)} & \colhead{} & \colhead{} & 
\colhead{} & \colhead{} & \colhead{(sec)}}
\startdata
PG 0923+201 & 09:25:54.7 & 19:54:04 & 0.190 & GHRS/G140L & z39r0106t & 22/11/96 & 18550 \\
PG 1309+355 & 13:12:17.7 & 35:15:21 & 0.184 & GHRS/G140L & z39r0206m & 26/07/96 &  2448 \\
PG 0804+761 & 08:10:58.6 & 76:02:43 & 0.100 & STIS/G140M & o4n301010 & 22/02/98 &  2448 \\
MRK 110 & 09:25:12.9 & 52:17:11 & 0.035 & STIS/G140M & o4n302010 & 20/05/99 & 2243 \\
PG 1049-005 & 10:51:51.5 & -00:51:18 & 0.3599 & STIS/G140L & o4n303010 & 05/11/98 & 150 \\
 & &  & & STIS/G140M & o4n303020 & & 1537 \\
PKS 1103-006 & 11:06:31.7 & -00:52:53 & 0.4233 & STIS/G140L & o4n304010 & 08/04/99 & 190 \\
           &  & & & STIS/G140M & o4n304020 & 08/04/99 & 1440 \\
TON 1542 & 12:32:03.6 & 20:09:29 & 0.064 & STIS/G140M & o4n305010 & 29/05/99 & 1990 \\
PG 1259+593 & 13:01:12.9 & 59:02:07 & 0.4778 & STIS/G140L & o4n307010 & 26/03/99 & 120 \\
  & &  & & STIS/G140M & o4n307020 & & 1808 \\
MRK 876 & 16:13:57.2 & 65:43:09 & 0.129 & STIS/G140M & o4n308010 & 19/09/98 & 2298 \\
\enddata
\end{deluxetable}

\clearpage
\begin{deluxetable}{cccccccccccc}
\setlength{\tabcolsep}{0.03in} 
\tabletypesize{\tiny}
\tablecaption{Target QSO-galaxies parameters and results\label{tab3}} 
\tablehead{
\colhead{QSO}  & \colhead{Galaxy} & \colhead{Type} & \colhead{M$_B$\tablenotemark{a}} & 
\colhead{Distance\tablenotemark{b}} & \colhead{$\rho$\tablenotemark{c}} & 
\colhead{$\rho / D_{25}$\tablenotemark{d}} & \colhead{V$_{gal}$\tablenotemark{e}} & 
\colhead{V$_{Ly\alpha }$\tablenotemark{f}} & \colhead{W\tablenotemark{g}} & 
\colhead{$b$\tablenotemark{h}} & \colhead{log N$_{HI}$\tablenotemark{i}} \\
\colhead{} & \colhead{} &\colhead{} & \colhead{} & \colhead{(Mpc)} & 
\colhead{(arcmin) (kpc)} & \colhead{} & \colhead{(\kms )} & \colhead{(\kms )} & 
\colhead{(\angst)} & \colhead{(\kms )} & \colhead{(cm$^{-2}$)}}
\startdata
PG 0923+201 & NGC 2903 & SB(s)d & -20.3 & 7.6 & 130.1  287 & 10.3 & 566 & - & $<0.14$ & - & - \\
PG 1309+355 & NGC 5033 Group & SA(s)c & -20.1 & 11.7 & 81.5  276 & 7.6 & 875 & 876.9 $\pm$ 18 & 1.10 $\pm$ 0.17 & 87, 147, 153 & 13.6, 14.0, 14.0 \\
PG 0804+761 & UGC 4238 & SBd & -18.7 & 20.6 & 22.7  136 & 9.5 & 1544 & 1569.9 $\pm$ 8 & 0.26 $\pm$ 0.03 & 131, 141 & 13.2, 13.2 \\
MRK 110 & NGC 2841 & SA(r)b & -21.2 & 14.1 & 84.0  344 & 10.4\tablenotemark{j} & 638 & - & $<0.09$ & - & - \\
PG 1049-005 & UGC 5985 & S0/a & -21.0 & 73.8 & 18.1  387 & 9.5 & 5538 & - & $<0.08$ & - & - \\
PKS 1103-006 & NGC 3521 & SAB(rs)bc & -20.9 & 10.7 & 51.9  162 & 4.7 & 805 & - & $<0.22$ & - & - \\
TON 1542 & UGC 7697 & Scd & -18.9 & 33.8 & 11.4  112 & 5.4 & 2536 & 2556.9 $\pm$ 12 & 0.29 $\pm$ 0.07 & 104 & 13.73 \\
PG 1259+593 & UGC 8146 & Scd & -16.7 & 8.9 & 21.3  55 & 6.1 & 669 & 679 $\pm$ 12 & 0.33 $\pm$ 0.08 & 109 & 13.78 \\
MRK 876 & NGC 6140 & SB(s)cd & -18.8 & 12.1 & 47.8  169 & 7.6 & 910 & 935 $\pm$ 5 & 0.39 $\pm$ 0.07 & 131 & 13.87 \\
\enddata
\tablenotetext{a}{Blue absolute magnitude of the galaxy, corrected for Galactic
and internal absorption, from the RC3.}
\tablenotetext{b}{Calculated using $H_O$=75 km/s/Mpc, except NGC2841 which has
a Cepheid distance.}
\tablenotetext{c}{Impact parameter between the QSO sightline and the center of the galaxy, in arcmin and in kpc}
\tablenotetext{d}{Ratio of the impact parameter to the optical diameter of the galaxy, 
measured at a surface brightness level of 25 mag/arc$^2$}
\tablenotetext{e}{Heliocentric velocity of the galaxy}
\tablenotetext{f}{Heliocentric velocity of the Ly$\alpha $ absorption. In the case of PG 1309+355 where the line is better fitted by three components, the
corresponding velocities are v$_1$=714.7, v$_2$=840.4 and v$_3$=1016.0 \kms .
Similarly for PG 0804+761 the two components have v$_1$=1506.0
and v$_2$=1614.5 \kms .}
\tablenotetext{g}{
Total equivalent width or 3$\sigma $ detection limit in case of non-detection}
\tablenotetext{h}{Doppler $b$ parameter of the line or different components of the line}
\tablenotetext{i}{HI column densities}
\tablenotetext{j}{There is another galaxy close to this sightline, UGC5047,
at cz=507 km/s. However it is a small dwarf and although it is only 108
$h^{-1}_{75}$ kpc away, its $\rho / D_{25}$ is over 35.}
\end{deluxetable}

\clearpage

\begin{deluxetable}{lccccccccccc} 
\setlength{\tabcolsep}{0.03in} 
\tabletypesize{\tiny}
\tablecaption{\label{tab4} Observing log for the VLA data and HI properties}
\tablehead{
\colhead{Source} & \colhead{Date} & \colhead{Array} & \colhead{t$_{int}$} & 
\colhead{Central Velocity} & \colhead{Beam\tablenotemark{a}} & 
\colhead{rms\tablenotemark{b}} & \colhead{V$_{sys}$\tablenotemark{c}} &
\colhead{$\Bigl<PA\Bigl>$\tablenotemark{d}} & 
\colhead{$\Bigl<i\Bigl>$\tablenotemark{e}} 
& \colhead{M$_{HI}$\tablenotemark{f}} & \colhead{D$_{HI}$\tablenotemark{g}} \\
\colhead{} & \colhead{} & \colhead{} & \colhead{(hours)} & \colhead{(\kms)} & 
\colhead{} & \colhead{(mJy beam$^{-1}$)} & \colhead{(\kms)} & 
\colhead{(\degrees )} & \colhead{(\degrees )} & \colhead{10$^9$ M$_{\odot}$} 
& \colhead{(arcmin)}} 
\startdata
 UGC4238 &  Nov 8, 1998 & VLA B/C & 1.5 & 1560 & $12"\times 12"$ & 2.4 & 1544 & 255 & 62 & 2.27 & 5.0 \\
 UGC7697 &  Apr 17, 2000 & VLA C & 4.6 & 2540 & $14"\times 14"$ & 1.0 & 2535 & 99 & 80 & 1.69 & 2.5 \\
 NGC6140 &  Apr 21, 2000 & VLA C & 4.3 & 910 & $14"\times 14"$ & 1.2 & 911 & 276 & 49 & 3.51 & 9.3 \\
\enddata
\tablenotetext{a}{Resulting beam size, full resolution (uniform weighting)}
\tablenotetext{b}{rms noise in channel maps at full resolution after Cleaning}
\tablenotetext{c}{Systemic (heliocentric) velocity of the galaxy}
\tablenotetext{d}{Average Position Angle of the HI distribution}
\tablenotetext{e}{Average Inclination of the HI distribution}
\tablenotetext{f}{Total Mass of HI, at the adopted distances listed in Table 2}
\tablenotetext{g}{HI diameter, at the 1 M$_\odot$pc$^{-2}$ level}
\end{deluxetable}

\clearpage
\begin{table}
\caption{ \label{tab5} Rotation Curves}
\begin{center}
\begin{tabular}{lccc}
Galaxy & Radius & Velocity & Error \\
 & (kpc) & (\kms) &  (\kms)  \\
\tableline
 UGC4238 &  & &  \\
 & 0.798 & 13.07 & 1.79 \\
 & 1.995 & 37.7 & 1.49 \\
 & 3.19 &  60.7 & 0.76 \\
 & 4.389 & 67.5 & 4.82 \\
 & 7.183 & 82.48 & 4.36 \\
 & 9.98 & 91.46 & 3.03 \\
 & 12.77 & 91.67 & 2.8 \\
 UGC7697 &  & &  \\
 & 2.13 & 58.77 &  0.52 \\ 
 & 4.42 & 73.99 &  2.87 \\ 
 & 6.72 & 93.10 &  0.82 \\
 & 9.01 & 105.52 & 1.07 \\
 & 13.60 & 107.3 & 2.25 \\
 NGC6140 &  & &  \\
 & 1.18 & 39.36 & 4.5 \\
 & 2.35 & 66.93 & 1.05 \\
 & 3.53 & 85.80 & 4.87 \\
 & 4.81 & 100.03 & 4.54 \\
 & 5.88 & 107.86 & 0.29 \\
 & 7.06 & 110.48 & 2.4 \\ 
 & 8.24 & 111.39 & 2.68 \\ 
 & 9.41 & 112.69 & 4.04 \\
 & 10.59 & 115.74 & 6.34 \\
 & 11.77 & 119.60 & 7.11 \\
 & 12.95 & 122.53 & 6.11 \\
 & 14.12 & 126.91 & 5.30 \\
 & 15.3 & 133.43 & 3.33 \\
 & 16.48 & 138.31 & 2.28 \\ 
 & 17.65 & 140.21 & 2.05 \\
 & 18.83 & 138.75 & 3.89 \\
 & 20.01 & 139.85 & 9.63 \\
\end{tabular}
\end{center}
\end{table}


\begin{thebibliography}{}
\bibitem []{a}Appleton, P.N., Foster, P.A., Davies, R.D. 1986, 221, 393
\bibitem []{b}Barcons, X., Lanzetta, K.M., Webb, J.K. 1995, Nature, 376, 321
\bibitem []{c}Barnes, D.G., Staveley-Smith, L., de Blok, W.J.G. et al 2001, MNRAS, 322, 486
\bibitem []{d}Begeman, K. 1987, Ph.D. thesis, Rijksuniversiteit Groningen
\bibitem []{e}Bland-Hawthorn, J., Freeman, K.C., Quinn, P.J. 1997, ApJ, 490, 143
\bibitem []{f}Bowen, D.V., Blades, J.C., Pettini, M. 1996, ApJ, 464, 141
\bibitem []{g}Bowen, D.V., Pettini, M., Blades, J.C. 2002, ApJ, 580, 169
\bibitem []{h}Carignan, C., C{\^o}t{\'e}, S., Freeman, K.C., Quinn, P.J. 1997, AJ, 113, 1585
\bibitem []{i}Chen, H-W., Lanzetta, K.M., Webb, J.K., Barcons, X. 1998, \apj, 498, 77
\bibitem []{j}Chen, H-W., Lanzetta, K.M., Webb, J.K., Barcons, X. 2001, \apj, 559, 654
\bibitem []{k}Cohen, R.J. 1979, MNRAS, 187, 838
\bibitem []{l}C{\^o}t{\'e}, S., Carignan, C., Freeman, K.C. 2000, \aj, 120, 3027
\bibitem []{m}C{\^o}t{\'e}, S., Taylor, A.R., Dewdney, P.E. 2002, ASP Conference Series, eds Taylor, Landecker, Willis, vol.276, p.92 
\bibitem []{n}Dav{\'e}, R., Hernquist, L., Katz, N., Weinberg, D. H. 1999, ApJ, 511, 521
\bibitem []{o}de Vaucouleurs, G., de Vaucouleurs, A., Corwin, J.R. et al 1991, ``Third
reference catalogue of bright galaxies", New York Springer Verlag (RC3)
\bibitem []{p}de Young, D.S., Heckman, T.M. 1994, ApJ, 431, 598
\bibitem []{az}Ebbets, D. 1995, ``Calibrating HST: Post Servicing Mission", eds A.Koratkar, C.Leitherer, p.207
\bibitem []{q}Dinshaw, N., Weymann, R.J., Impey, C.D. et al 1997, ApJ, 491, 45
\bibitem []{r}Dinshaw, N., Foltz, C.B., Impey, C.D., Weymann, R.J. 1998, ApJ, 494, 567
\bibitem []{s}Gibson, B.K., Giroux, M.L., Penton, S.V. et al  2001, AJ, 122, 3280
\bibitem []{t}Hartmann, D., Burton, W.B. 1997, ``Atlas of galactic neutral hydrogen", 
Cambridge University Press
\bibitem []{u}Hoffman, G.L., Lu, N.Y., Salpeter, E.E., Connell, B.M., Fromhold-Treu, R. 1998, ApJ, 500, 789
\bibitem []{v}Huchtmeier, W., Richter, O. 1989, ``A General Catalog of HI observations of Galaxies", New York Springer.
\bibitem []{w}Kerr, F.J., de Vaucouleurs, G. 1955, AuJPh, 8, 508
\bibitem []{x}Lanzetta, K.M., Bowen, D.V., Tytler, D., Webb, J.K. 1995, ApJ, 442, 538
\bibitem []{y}Lockman, F.J., Savage, B.D. 1995, ApJS, 97, 1
\bibitem []{z}McLin, K.M., Stocke, J.T., Weymann, R.J., Penton, S.V., Shull, J.M. 2002, ApJ, 574, L115
\bibitem []{aa}Mac Low, M.M., McCray, R. 1988, ApJ, 324, 776
\bibitem []{ab}Macri, L.M., Stetson, P.B., Bothun, G.D. et al 2001, ApJ, 559, 243
\bibitem []{ac}Maloney, P. 1993, ApJ, 414, 41
\bibitem []{ad}Meurer, G.R., Carignan, C., Beaulieu, S.F., Freeman, K.C. 1996, 111, 1551
\bibitem []{ae}Miralda-Escud{\'e}, J., Cen, R., Ostriker, J.P., Rauch, M. 1996, ApJ, 471, 582
\bibitem []{af}Morris, S.L., Jannuzi, B.T., Weymann, R.J. 2002, ASP Conference Series, vol. 254, eds. J.S.Mulchaev \& J.Stocke, p.72
\bibitem []{ag}Nath, B.B., Trentham, N. 1997, MNRAS, 291, 505
\bibitem []{ah}Navarro, J., Frenk, C., White, S.D.M. 1996, \apj, 462, 563
\bibitem []{ai}Penton, S.V., Stocke, J.T., Shull, J.M. 2002, ApJ, 5
\bibitem []{aj}Persic, M., Salucci, P., Stel, F. 1996, MNRAS, 281, 27
\bibitem []{ak}Rhee, M.H., van Albada, T.S. 1996, A\&AS, 115, 407
\bibitem []{al}Richter, P., Savage, B.D., Wakker, B.P. et al 2001, ApJ, 549, 281
\bibitem []{am}Shapiro, P. \& Field, G. 1976, ApJ, 205, 762
\bibitem []{an}Shull, J.M., Penton, S.V., Stocke, J.T. et al 1998, AJ, 116, 2094
\bibitem []{ao}Shull, J.M., Giroux, M.L., Penton, S.V. et al 2000, ApJ, 538, L13
\bibitem []{ap}Silk, J., Sunyaev, R.A. 1976, Nature, 260, 508
\bibitem []{aq}Smith, D.R., Bernstein, G.M., Fischer, P., Jarvis, M. 2001, ApJ, 551, 643
\bibitem []{ar}Steidel, C.C., Kollmeier, J.A., Shapley, A.E. et al 2002, ApJ, 570, 526
\bibitem []{as}Swaters, R.A., Sancisi, R. \& van der Hulst, J.M. 1997,  \aj, 491, 140 
\bibitem []{at}Tripp, T.M., Lu, L., Savage, B.D. 1998, \apj, 508, 200
\bibitem []{au}van Gorkom, J., Carilli, C., Stocke, J., Perlman, E. \& Shull, J.M. 1996, AJ, 112, 1397 
\bibitem []{av}van Gorkom, J.H. 1991, in "Atoms, Ions, and Molecules: New Results in Spectral Line Astrophysics", eds. Haschick and Ho, p.1
\bibitem []{aw}V\'eron-Cetty, M., V\'eron, P. 1996, {``A Catalogue of Quasars and
 Active Nuclei"}, ESO Scientific Report No.17
\bibitem []{ax}Wakker, B.P. 1991, A\&AS, 90, 495
\bibitem []{ay}Wakker, B.P., Kalberla, P.M.W., van Woerden, H. et al 2001, ApJS, 136, 537
\end{thebibliography}
\end{document}